\documentclass[aip,amsmath,amssymb,preprint]{revtex4-1}
\usepackage[letterpaper, portrait, margin=1in]{geometry}
\usepackage{graphicx}
\usepackage[version=4]{mhchem}
\usepackage{xurl}
\usepackage{bm}
\usepackage{bbm}

\usepackage{jabbrv}

\usepackage{wrapfig}

\usepackage{hyperref}
\usepackage{xr} 
\hypersetup{colorlinks=true,urlcolor=blue,citecolor=blue,linkcolor=blue}

\begin{document}

\title{Dynamic Ensembles of Phosphine-Stabilized Gold Nanoclusters}
\author{Caitlin A. McCandler$^{*}$}
\address{Initiative for Computational Catalysis, Flatiron Institute, New York, NY 10010, USA}
\email{cmccandler@flatironinstitute.org}
\author{Disha Sanwal}
\address{University of Chicago, Chicago, IL 60637, USA}
\author{Jutta Rogal}
\address{Initiative for Computational Catalysis, Flatiron Institute, New York, NY 10010, USA}

\date{\today}

\begin{abstract}
Atomically precise phosphine-stabilized gold nanoclusters are commonly characterized by single-crystal X-ray diffraction, yet the extent to which these static structures represent finite-temperature behavior remains unclear. 
To explore the free-energy landscapes, equilibrium populations, and isomerization kinetics of these nanoclusters in the gas phase, we establish a general framework that combines molecular dynamics simulations based on a machine-learned interatomic potential with Markov state models (MSMs). 
Analysis of the MSMs indicates that experimentally reported crystal structures frequently correspond to minor metastable states or transient configurations rather than the dominant finite-temperature structures. Increasing ligand coverage systematically alters both the thermodynamics and kinetics of structural rearrangements, driving the transition from planar to three-dimensional gold cores while accelerating isomerization dynamics. 
Moreover, catalytically accessible geometries are often only minor members of the equilibrium ensemble, highlighting a trade-off between structural stability and surface accessibility. 
These results emphasize that ligand-protected nanoclusters need to be viewed as dynamic ensembles and their finite-temperature behavior cannot be fully captured by their corresponding crystallographic structures alone. 
\end{abstract}

\maketitle

Understanding the structure–property relationships of monolayer-protected metal nanoclusters remains a central challenge in nanoscience, especially in solution-phase environments relevant to catalysis and sensing.\cite{matus_understanding_2023,yao_molecule-like_2025} Phosphine-stabilized gold nanoclusters are one such class of nanoclusters which have specific properties related to their size and structure. Their distinct geometries can be resolved with atomic precision, most commonly through single-crystal X-ray diffraction. While such crystallographic approaches have been important in providing precise atomic structures, they inherently capture static snapshots of the nanocluster in the crystal that cannot fully reflect the dynamic behavior of clusters under operating conditions in solution or in the gas phase. 
In these nanocluster crystals, 
the observed geometries are stabilized not only 
by intramolecular interactions within the ligand shell and metal–ligand framework 
but also by crystal packing.\cite{corpinot_practical_2019,desiraju_crystal_2013} 


Recent experimental observations suggest that distinct nanocluster structures, initially isolated and characterized as single crystals, can interconvert or decay into different conformations in solution.\cite{deng_structural_2024} This raises fundamental questions about the persistence of atomically precise geometries and the extent to which structures characterized in the crystalline phase are relevant in solution. Such considerations are particularly important in applications where function is tightly coupled to geometry. In catalysis, for example, identifying cluster geometries that resist agglomeration while maintaining accessible and reactive surface sites remains a key objective.\cite{du_atomically_2020} Additionally, for sensing and photophysical applications, the structural rigidity of a cluster is related to its quantum yield, as increased vibrational degrees of freedom are known to promote nonradiative relaxation pathways.\cite{zhong_suppression_2023,li_photoluminescence_2026}

To explore the finite-temperature behavior of nanoclusters beyond their static crystallographically determined structures, computational approaches capable of accessing both atomistic detail and long-timescale dynamics are required. Here, we combine molecular dynamics (MD) simulations driven by machine-learned interatomic potentials (MLIPs) with Markov state models (MSMs)\cite{husic_markov_2018,prinz_markov_2011,wang_constructing_2018} to characterize the finite-temperature free-energy landscapes and isomerization dynamics of phosphine-stabilized gold nanoclusters. Recent advances in MLIPs have made it possible to obtain accurate energetics for atomically precise nanoclusters, enabling MD studies of thermal fluctuations, chiral inversion, ligand dynamics, and intercluster reactions in thiolate- and thioformimidate-stabilized gold clusters.\cite{mccandler_goldthiolate_2024,sabooni_asre_hazer_thermal_2025,sabooni_asre_hazer_machine_2026,tiwari_atomistic_2025} 
Here, we focus on phosphine-stabilized gold clusters which are of particular interest for catalysis due to the weaker binding of phosphine ligands to gold compared with other common ligands, such as thiols.\cite{kawawaki_toward_2021}
By fitting an MLIP for this system, MD simulations with near first-principles accuracy on timescales reaching the microsecond regime become computationally tractable, providing access to rare structural transitions that cannot be observed with conventional \textit{ab initio} MD. 
The resulting dynamical trajectories are analyzed using MSMs, a coarse-graining framework developed extensively in the biophysics community for describing conformational dynamics and long-timescale kinetics~\cite{husic_markov_2018,prinz_markov_2011,wang_constructing_2018} that we recently extended to application in reactive systems.~\cite{mccandler_markov_2026} By introducing a structural featurization tailored to metal nanoclusters, we demonstrate that MSMs can further be adapted to the study of atomically precise nanoclusters. This framework enables the identification of metastable states, quantification of isomerization pathways, and connection of atomically resolved dynamics to ensemble-level behavior.

The analysis of the MSMs reveals that structures stabilized and characterized in the solid state are often not the most stable ones in the gas phase. Furthermore, the number of ligands significantly impacts the structural diversity and isomerization kinetics. The implications of these dynamic equilibria for catalysis are investigated by relating entropically favored geometries to the accessibility of the gold core to adsorbates. Together, this work demonstrates the need to move beyond static structural descriptions toward a dynamic, ensemble-based understanding of nanocluster structures and behavior.

\section*{Dynamical Maps of \ce{A\lowercase{\text{u}}_{\lowercase{\text{n}}}L_{\lowercase{\text{m}}}} Ensembles}

The dynamical stability, free-energy landscapes, equilibrium isomer populations, and isomerization kinetics of gold nanoclusters are investigated for experimentally reported \ce{Au_{n}(PR3)_{m}} ($n\leq$12, $m\leq n$) clusters listed in the Cambridge Structural Database (CSD)\cite{groom_cambridge_2016}, including \ce{Au4L4},\cite{yang_new_1994,zeller_tetrahedral_1993} \ce{Au6L6},\cite{briant_synthesis_1986,bellon_octahedral_1973} \ce{Au7L6},\cite{susukida_kohei_ccdc_2020} \ce{Au7L7},\cite{schulz-dobrick_ccdc_2008,van_der_velden_intermediates_1984,marsh_crystal_1984,marsh_thoughts_1995} \ce{Au8L6},\cite{yang_new_1994} \ce{Au8L7},\cite{van_der_velden_synthesis_1981,van_der_velden_reactions_1983} \ce{Au8L8},\cite{schulz-dobrick_m_ccdc_2009} \ce{Au9L8},\cite{shen_hui_ccdc_2020,schulz-dobrick_ccdc_2007,malinina_ccdc_2008,smits_x-ray_1983,schulz-dobrick_ccdc_2009-1,schulz-dobrick_ccdc_2009-2,schulz-dobrick_ccdc_2007-1,schulz-dobrick_ccdc_2009,wang_ccdc_2019-1,wang_ccdc_2019} and \ce{Au11L10}.\cite{copley_novel_1996} Throughout this work, the experimentally used phosphine ligands are represented by trimethylphosphine (\ce{P(CH3)3}), which provides a more realistic approximation to bulky \ce{PR3} ligands than the commonly used \ce{PH3} model. In particular, \ce{PH3} binds more weakly than bulkier phosphines,\cite{parrish_role_2019} and in preliminary finite-temperature simulations this led to unrealistically frequent ligand dissociation. Trimethylphosphine therefore provides a compromise between retaining important ligand steric and binding effects and maintaining computational tractability.

\begin{figure}
    \centering
    \includegraphics[width=\linewidth]{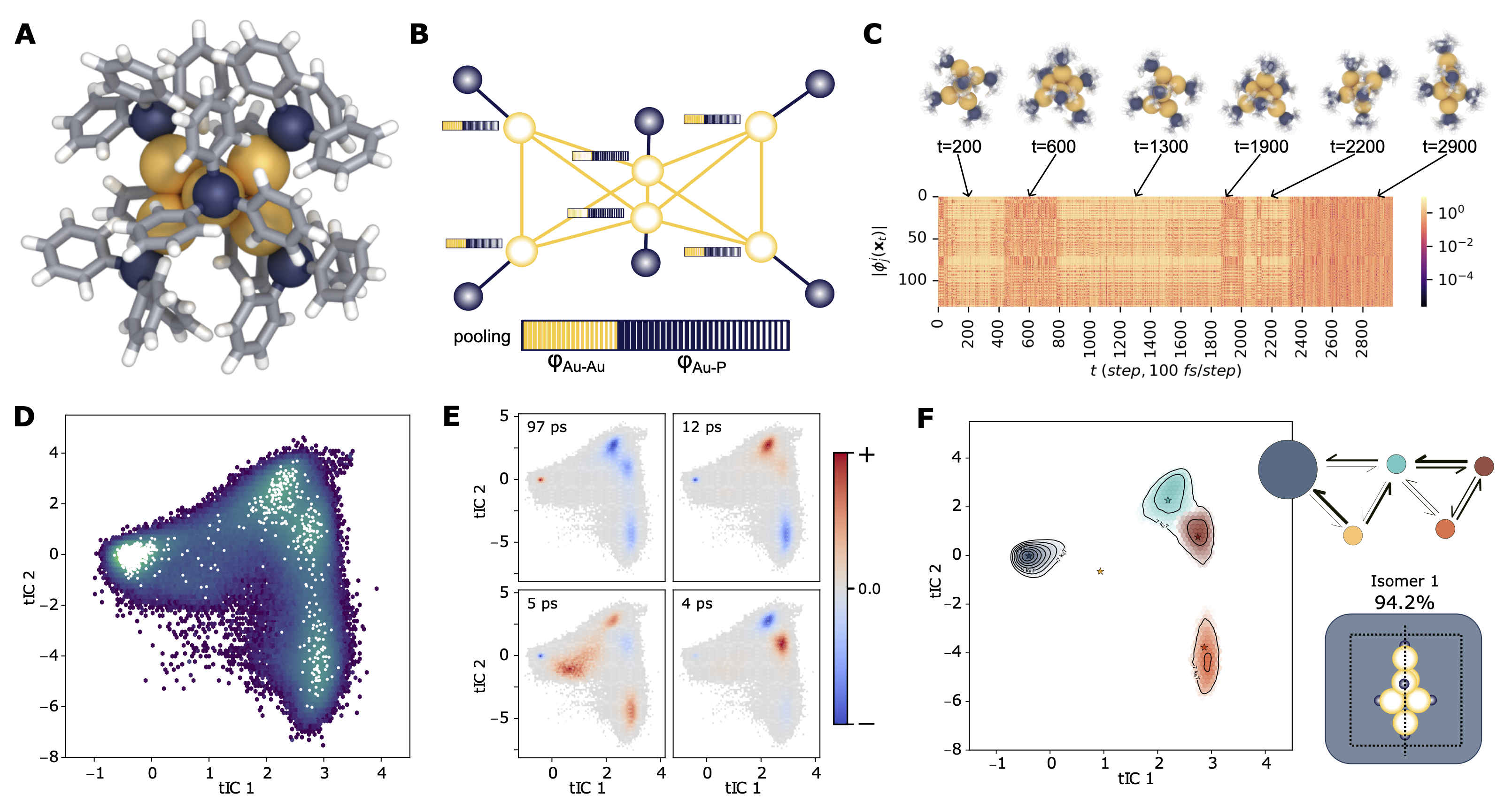}
    \caption{Overview of the computational workflow used to construct dynamical maps of phosphine-stabilized gold nanoclusters. A. Crystal structure of an \ce{Au6(PPh3)6} cluster.\cite{briant_synthesis_1986} B. Feature vectors are calculated for each snapshot in the MD trajectory as ACE basis functions pooled over all Au atoms. C. Isomerization events are tracked by changes in the feature vectors over the simulation trajectories. The structure features reveal both large and small geometry fluctuations. 
    D. The feature vectors representing each simulated frame are projected into TICA space and K-means clustering groups structures into discrete state assignments. The density of states and cluster centers are plotted in the first two dimensions of TICA space. E. The first eigenmode of the transition matrix between discrete states reveals the slowest dynamical mode in this system. F. States that interconvert relatively rapidly are grouped into a single isomer. Coarse-grained dynamical models provide the frequency of isomerization.}
    \label{fig:msmfitting}
\end{figure}

To characterize the free-energy landscapes, equilibrium isomer populations, and isomerization kinetics of each system, MLIP-based MD simulations were performed in the gas phase at $T=300$~K, collecting a total of $288-1536$~ns of trajectory data for the different systems. Details regarding the MD simulations and MLIP training can be found in the \nameref{sec:methods} section. 
The resulting trajectories were used to construct MSMs following the workflow illustrated in Figure~\ref{fig:msmfitting}: for each simulation frame, the configuration is represented by a feature vector, comprising a set of structural descriptors based on atomic cluster expansion (ACE) functions\cite{drautz_atomic_2019} (Fig.~\ref{fig:msmfitting}B and C), that are projected into a low-dimensional space using time-lagged independent component analysis (TICA)\cite{molgedey_separation_1994,perez-hernandez_identification_2013} (Fig.~\ref{fig:msmfitting} D).  Within the TICA space, configurations are clustered into microstates, which are then coarse-grained into metastable macrostates based on dynamical information. Throughout this work, 
we use the macrostates to identify different \textit{isomers}, 
defined as an ensemble of structurally similar configurations connected by rapid fluctuations relative to the slow transitions between distinct metastable states. 
Details regarding the MSM construction are provided in the \nameref{sec:methods} section.

\begin{figure}
    \centering
    \includegraphics[width=\linewidth]{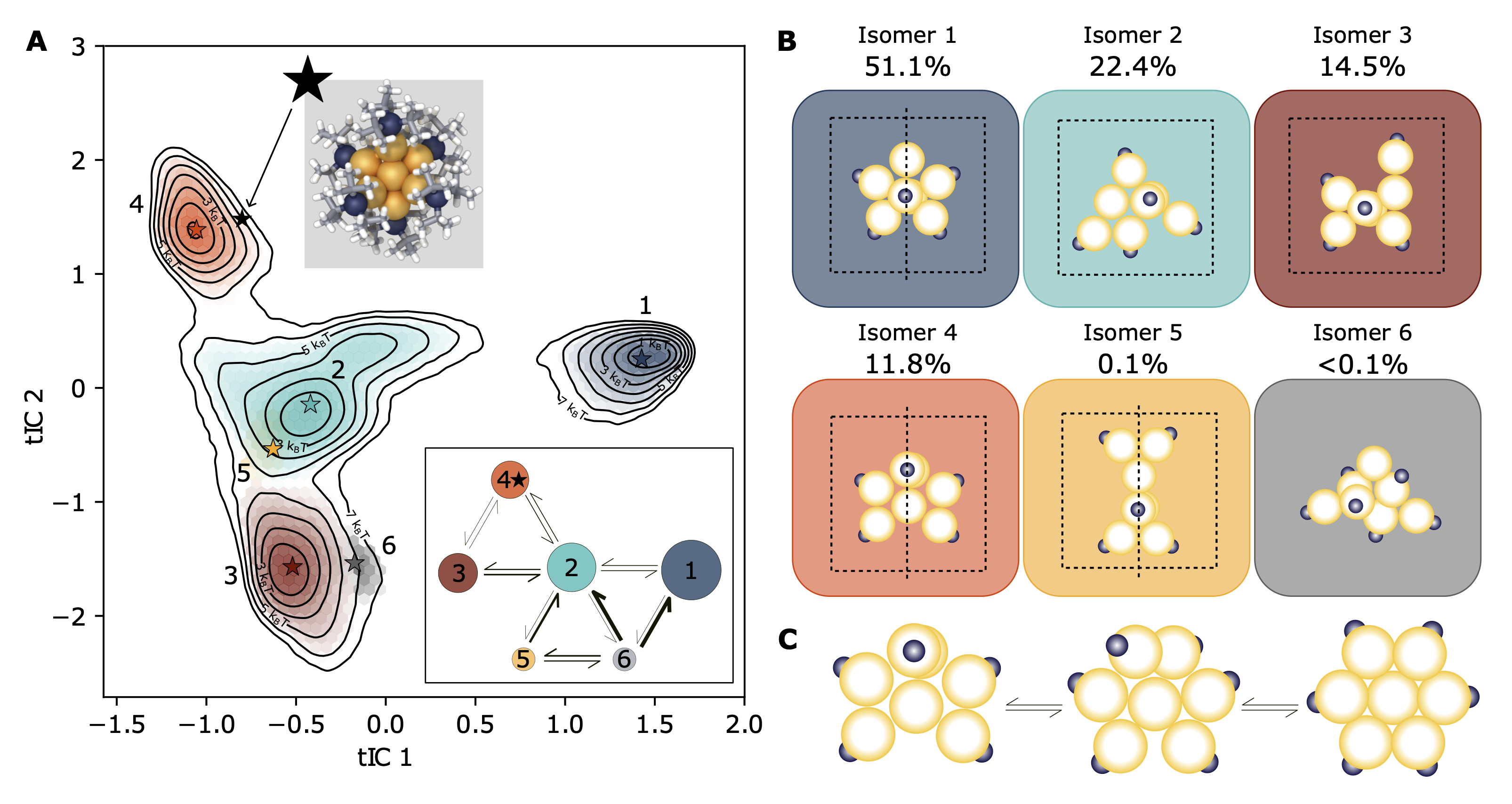}
    \caption{A dynamical picture of the isomerization of \ce{Au7L6} at 300~K in the gas phase. A. The free energy 
    projected onto the first two dimensions of TICA space, with an inset schematic of the transition likelihood between each isomer. The crystal structure of the known \ce{Au7L6} cluster\cite{susukida_kohei_ccdc_2020} is projected onto this map (black star) and is found to correspond to a minor isomer (isomer 4, 11.8\%). B. The equilibrium populations and representative structures of each isomer. Snapshots corresponding to structures at the center of each free-energy basin (stars on map) are shown together with schematics indicating the mirror symmetries of each isomer. Au atoms are denoted by large yellow spheres and P ligands are denoted by small blue spheres. C. The mechanism of fast fluctuations between the most representative structure of isomer 4 and the known crystal geometry of \ce{Au7L6}.}
    \label{fig:map_au7l6}
\end{figure}


The MSM analysis is discussed in detail for \ce{Au7L6} as a representative example, with analogous analyses for the remaining cluster sizes reported in the SI (Figs.~\ref{fig:Au4L4}-\ref{fig:Au11L10}). To provide a low-dimensional representation of the slow structural rearrangements observed in the simulations, the free-energy landscape of \ce{Au7L6} is projected onto the first two TICA components (4 dimensions in total, see SI Sec. \ref{subsec:MSMfitting}), as shown in Fig.~\ref{fig:map_au7l6}A. 
The metastable macrostates obtained from the dynamical clustering of the MSM microstates are generally correlated with the free-energy basins in this projection. 
For each isomer, the structure closest in TICA space to the centroid of the free-energy basin is selected as a representative example, marked by a star in Fig.~\ref{fig:map_au7l6}A and schematically shown in Fig.~\ref{fig:map_au7l6}B.
The free-energy contours and equilibrium populations of each isomer are reweighted according to the static populations of each Markov state (details in the \nameref{sec:methods} section). The inset summarizes the dominant transition pathways between macrostates occurring 
on a timescale of $\Delta t = 100$~fs.
For quantitative transition probabilities, see SI Fig.~\ref{fig:Au7L6}. For \ce{Au7L6}, isomer 2 has a non-zero transition probability to all other states within $\Delta t = 100$~fs, meaning it may act as an intermediate in isomerization between less similar geometries.

The experimentally determined crystal structure of \ce{Au7L6} has a puckered hexagonal geometry with three-fold symmetry. When projected onto the simulated free-energy landscape, this structure falls within isomer 4, a metastable state comprising only 11.8\% of the equilibrium ensemble. Interestingly, the representative structure of isomer 4 does not possess the same hexagonal geometry as the crystal structure, indicating that the experimentally measured nanocluster geometry represents only one configuration within a broader metastable basin. 
Because these geometries 
fall into the same metastable state in the TICA projection at a lag time of 0.5~ps, 
their interconversion is best understood as a relatively fast fluctuation rather than a slow isomerization event. By comparison, the fastest resolved isomerization process in this system has a characteristic timescale of 43~ps. The corresponding fluctuation mechanism, depicted schematically in Fig.~\ref{fig:map_au7l6}C, was also directly observed in the MD simulation. 

The experimentally reported crystal structures of \ce{Au4L4}, \ce{Au6L6}, \ce{Au7L7}, and \ce{Au8L7} also correspond to only minor isomers with equilibrium populations of 0.2\%, 2.9\%, 12.7\%, and 7.7\%, respectively (SI Figs.~\ref{fig:Au4L4}, \ref{fig:Au6L6}, \ref{fig:Au7L7}, \ref{fig:Au8L7}), and the reported crystal structures of \ce{Au8L6} and \ce{Au8L8} do not coincide with metastable basins in the simulated free-energy landscapes at all (SI Figs.~\ref{fig:Au8L6} and \ref{fig:Au8L8}). The \ce{Au9L8} and \ce{Au11L10} crystal structures, however, do correspond to the dominant equilibrium isomers, accounting for 94.6\% and 99.0\% of the equilibrium populations, respectively (SI Figs.~\ref{fig:Au9L8} and \ref{fig:Au11L10}). Aside from the very small \ce{Au4L4} cluster, each system size has at least 4 isomers that are structurally and kinetically distinct from one another (SI Table~\ref{tbl:fits}). Collectively, these dynamical maps show that phosphine-stabilized gold nanoclusters should be viewed as equilibrium ensembles of interconverting isomers rather than as single static structures. 
The crystallographic geometries do not necessarily correspond to dominant equilibrium states but may represent minor metastable states or transient configurations within larger free-energy basins, depending on the cluster size. By simultaneously resolving equilibrium populations, interconversion pathways, and transition kinetics, the MSMs provide a quantitative framework for relating experimentally observed crystal structures to the underlying free-energy landscape.

\section*{Impact of Ligation on Dynamical Properties}
\phantomsection
\makeatletter
\def\@currentlabelname{Impact of Ligation on Dynamical Properties}
\makeatother
\label{sec:ligation}

\begin{figure}
    \centering
    \includegraphics[width=\linewidth]{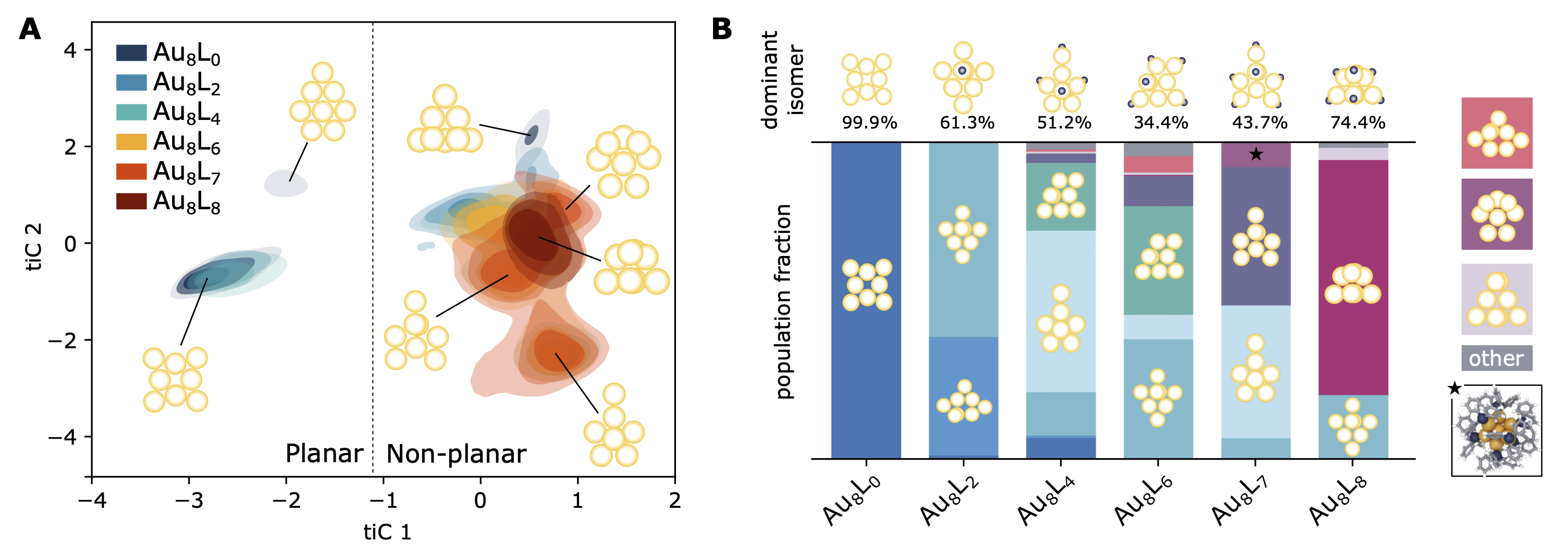}
    \caption{Effect of ligand coverage on the structural dynamics of \ce{Au8} nanoclusters. A. Free energy surfaces of \ce{Au_{8}L_{m}} (m = 0, 2, 4, 6, 7, and 8) projected onto the first two TICA components  of Au-only interactions across all \ce{Au8L_{m}} simulations, illustrating the evolution of the gold core's configurational landscape with increasing ligand coverage. The bare \ce{Au8} cluster predominantly samples planar geometries, whereas the addition of ligands stabilizes non-planar structures. Representative gold core geometries are shown for the major free energy basins. Separate plots for each cluster are included in SI Figure~\ref{fig:au8breakout}. B. Population distributions of the Au cores in the equilibrium ensembles of \ce{Au_{8}L_{m}} clusters, and the most dominant isomer as a function of the number of protecting ligands, $m$. The experimentally reported \ce{Au8L7}\cite{van_der_velden_synthesis_1981,van_der_velden_reactions_1983} crystal structure (black star) appears only as a minor isomer in the \ce{Au8L7} ensemble and the gold core geometries of reported \ce{Au8L6}\cite{yang_new_1994} and \ce{Au8L8}\cite{schulz-dobrick_m_ccdc_2009} crystals are not represented among the observed isomers.
}
    \label{fig:au8}
\end{figure}

As ligand coverage is a primary determinant of nanocluster stability and reactivity, we investigate how the number of ligands modifies the dynamical stability of gold nanoclusters. Using \ce{Au8} as a representative example, we examine the equilibrium ensembles and conformational dynamics of \ce{Au8L_m} ($m=0, 2, 4, 6, 7, 8$). 
Additional simulations were run for structures with $m=0, 2, 4$ that were not included in the previous section and MSMs were fitted to their dynamics (details in SI Table~\ref{tbl:fits}). 
The dynamical maps and isomer populations for these additional systems are included in the SI (Figs.~\ref{fig:Au8L0}-\ref{fig:Au8L4}). The effect of ligand coverage on the free-energy landscape is summarized in Figure \ref{fig:au8}A. Projecting the equilibrium ensembles into a common TICA space (fitted to only Au-Au structural basis functions from trajectory data of every \ce{Au8L_{m}} system) shows the structural evolution of the Au core as a function of ligation.  The bare \ce{Au8} cluster resides almost exclusively ($\sim$99.9\%) in a single free-energy basin corresponding to a planar geometry with four-fold symmetry. Upon addition of only two ligands, non-planar structures emerge as dominant states, and increasing ligand coverage progressively shifts the equilibrium population toward three-dimensional geometries, with the fully ligated \ce{Au8L8} dominant structure adopting the most compact gold-core geometry observed across the series. This behavior is consistent with previous electronic structure calculations that show that bare Au, stabilized by strong relativistic effects, has planar geometries even at large sizes (\ce{Au7}-\ce{Au13} depending on the computational method)\cite{hakkinen_gold_2000,fernandez_trends_2004,xiao_planar_2004,walker_structure_2005,dong_global_2007,sekhar_de_understanding_2010,assadollahzadehSystematicSearchMinimum2009,fa_bulk_2005,li_size_2007,johansson_at_2014,kinaci_unraveling_2016,goldsmith_two--three_2019,chaves_evolution_2017,manna_database_2023,wu_nonlocal_2019} and a previous computational study of phosphine-stabilized gold clusters which found ligation to stabilize 3D geometries in Au clusters.\cite{mccandler_phosphine-stabilized_2023} Clusters with similar ligand coverages preferentially sample similar regions of configuration space, indicating that the preferred gold-core geometries evolve continuously as ligands are added. Some core geometries, however, are stabilized only at specific ligand coverages because they possess symmetries that can be realized only for particular ligand counts.

The equilibrium populations of the resulting macrostates are summarized in Figure \ref{fig:au8}B. The bare cluster is almost exclusively found in a single isomer (99.9\%), while the fully ligated \ce{Au8L8} cluster is similarly dominated by one structure (74.4\%). Between these two extremes, however, the structural diversity increases substantially. It reaches a maximum near half coverage, where \ce{Au8L6} exhibits twenty distinct metastable isomers, eight of which contribute more than 2.5\% to the equilibrium ensemble population. The broad distribution of populations indicates that intermediate ligand coverages produce relatively shallow free-energy landscapes containing many competing minima of comparable stability. In contrast, both the bare and fully protected clusters occupy deep, well-defined basins corresponding to a single dominant structure. The experimentally determined crystal structure of \ce{Au8L7}\cite{van_der_velden_synthesis_1981,van_der_velden_reactions_1983} is also represented in the dynamic ensemble (Figure \ref{fig:au8}B), although it appears only as a minor isomer. Conversely, the reported crystal structures of \ce{Au8L6}\cite{yang_new_1994} and \ce{Au8L8}\cite{schulz-dobrick_m_ccdc_2009} have gold-core geometries that are not sampled among the major equilibrium isomers identified here. These observations emphasize that experimentally observed crystal structures do not necessarily correspond to the dominant gas-phase structures and highlight the importance of explicitly accounting for conformational dynamics when interpreting nanocluster geometries.

Ligand coverage also has a large effect on the isomerization kinetics. The slowest implied timescales are on the order of $10^4$~ps for the sparsely ligated clusters (\ce{Au8L0}-\ce{Au8L4}), but decrease to $10^3$-$10^2$~ps at higher ligand coverages (\ce{Au8L6}-\ce{Au8L8}; see SI Table~\ref{tbl:fits}). Although intermediate ligand coverages exhibit the greatest number of metastable structures, these states interconvert much more rapidly once the cluster becomes heavily ligated. In contrast, the sparsely ligated clusters remain trapped for much longer times within individual free-energy basins. 
Overall, ligand coordination not only modifies the thermodynamic landscape but also lowers kinetic barriers separating structurally distinct gold-core conformations, enhancing isomerization kinetics.

\section*{\ce{Au9(PR3)8} Polymorphs}
\begin{figure}
    \centering
    \includegraphics[width=\linewidth]{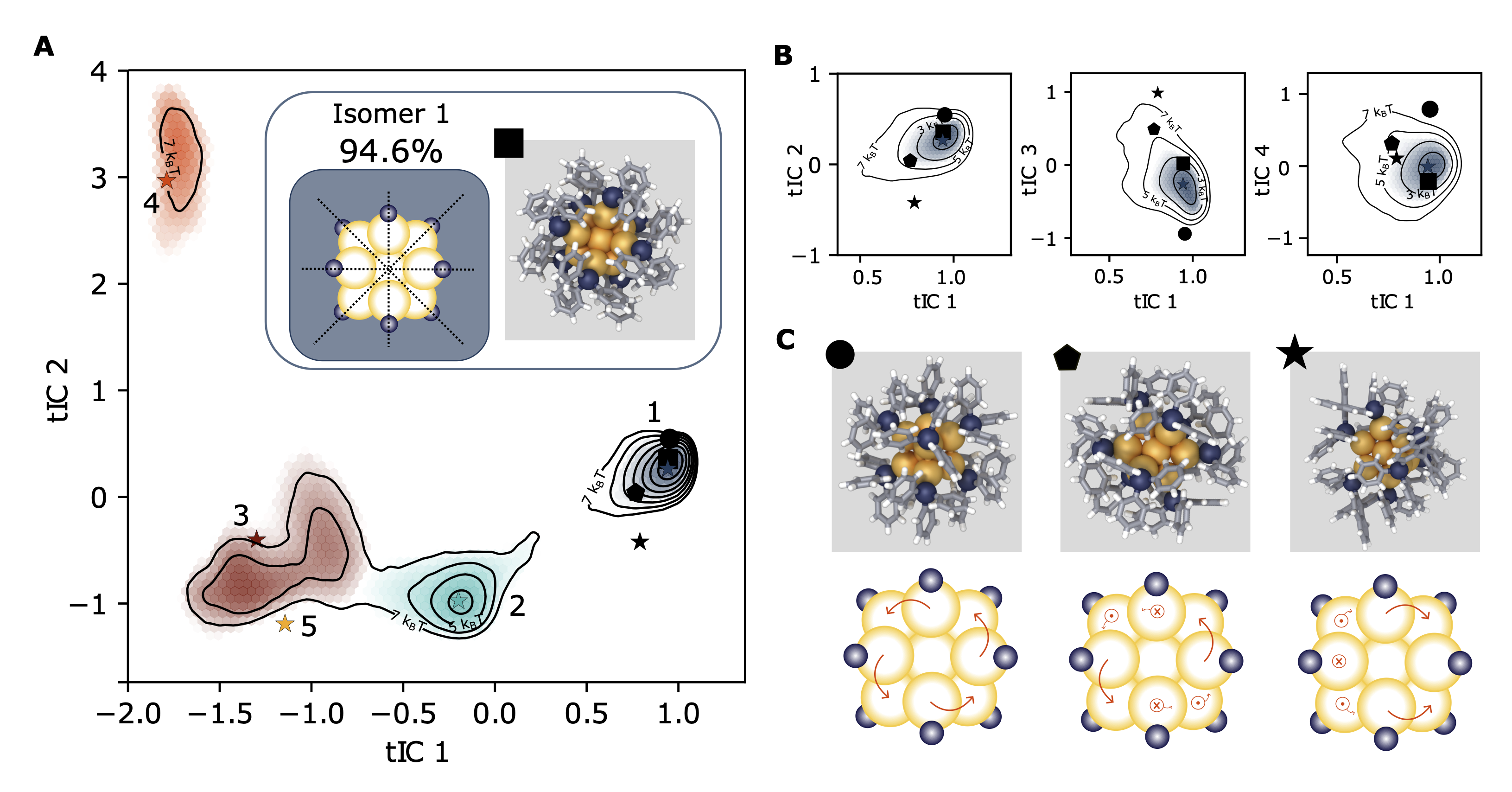}
    \caption{Multiple experimentally observed crystal geometries of \ce{Au9L8}\cite{shen_hui_ccdc_2020,schulz-dobrick_ccdc_2007,malinina_ccdc_2008,smits_x-ray_1983,schulz-dobrick_ccdc_2009-1,schulz-dobrick_ccdc_2009-2,schulz-dobrick_ccdc_2007-1,schulz-dobrick_ccdc_2009,wang_ccdc_2019-1,wang_ccdc_2019} correspond to a single gas-phase isomer. A. Free-energy surface projected onto TICA space, where the four known crystal structures (black square, circle, pentagon and star) all lie within a single free energy basin and rapidly interconvert, indicating that they are configurations of the same isomer. The representative structure of this isomer (isomer 1) and the crystal structure most similar to it (square marker) are shown in the inset. B. Additional projections onto higher TICA coordinates reveal that the crystal structure denoted by the square marker lies closest to the most representative structure of isomer 1, while the remaining crystal geometries occupy nearby regions within the same basin. The crystal structure denoted by the star marker has bidentate ligands and is therefore not directly comparable to the monodentate structures, although it remains within the same free-energy basin. C. Schematic illustration of collective Au rearrangements required to transform the most representative structure into the alternative crystal geometries. In the schematics, $\odot$ and $\otimes$ indicate vectors pointing out of and into the page, respectively.}
    \label{fig:au9_polymorphs}
\end{figure}

The \ce{Au9(PR3)8} system provides a particularly interesting example of how crystallographically distinct structures do not necessarily correspond to separate finite-temperature isomers. Four different crystal geometries have been reported for phosphine-protected \ce{Au9(PR3)8} clusters, which differ only in the arrangement of phosphine-bound gold atoms around a central gold core atom.\cite{shen_hui_ccdc_2020,schulz-dobrick_ccdc_2007,malinina_ccdc_2008,smits_x-ray_1983,schulz-dobrick_ccdc_2009-1,schulz-dobrick_ccdc_2009-2,schulz-dobrick_ccdc_2007-1,schulz-dobrick_ccdc_2009,wang_ccdc_2019-1,wang_ccdc_2019} When these experimentally determined geometries are projected onto the TICA free-energy landscape obtained from our MD simulations, they all fall within the same free-energy basin (Fig.~\ref{fig:au9_polymorphs}A) which corresponds to the most populated isomer (isomer 1, 94.6\%). As the TICA space for this system is 7-dimensional, we examine projections onto higher TICA dimensions (Fig.~\ref{fig:au9_polymorphs}B), confirming that the four experimentally observed geometries remain localized within the same free-energy basin. 
Therefore, the geometries of the reported crystal structures are connected by rapid thermal fluctuations rather than occupying distinct metastable states separated by kinetic barriers.

The crystal structure closest to the most representative geometry of this isomer is indicated by the square marker in Fig.~\ref{fig:au9_polymorphs}A and has the highest symmetry (4-fold rotational symmetry). Notably, one of the crystal structures (star marker) was synthesized with bidentate ligands experimentally, and while it also resides within the same basin, it is the furthest from the most representative structure, which is likely due to the additional constraints on the structure coming from the ligand-ligand bonding. The structural differences between each of these crystal geometries arise only from subtle rearrangements of the outer phosphine-bound gold atoms. Representative fluctuation mechanisms are illustrated schematically in Fig.~\ref{fig:au9_polymorphs}C showing that transformations are easily accessible with only small rearrangements of the Au core. 
Consequently, crystallographically resolved geometries should be interpreted as snapshots of a single dynamic gas-phase isomer rather than as distinct molecular species.

\section*{Catalytically active geometries}

The relationship between structural stability and catalytic accessibility presents an inherent trade-off in ligand-protected nanoclusters. As described in the section \nameref{sec:ligation}, increasing ligand coverage stabilizes the gold core by shielding the surface from the surrounding environment. Consequently, highly protected species such as \ce{Au4L4}, \ce{Au6L6}, \ce{Au7L7}, \ce{Au8L8}, \ce{Au9L8}, and \ce{Au11L10} are commonly isolated experimentally as stable nanoclusters. Although \ce{Au9L8} and \ce{Au11L10} each contain one fewer ligand than gold atoms, the unligated gold atom preferentially occupies the core of the cluster, leaving the external surface effectively fully protected. While such extensive ligand coverage inhibits agglomeration and enhances structural stability, it also sterically limits access of reactants to the gold surface. As a result, the geometries that are most stable are not necessarily the most accessible for catalytic reactions.

\begin{figure}
    \centering
    \includegraphics[width=\linewidth]{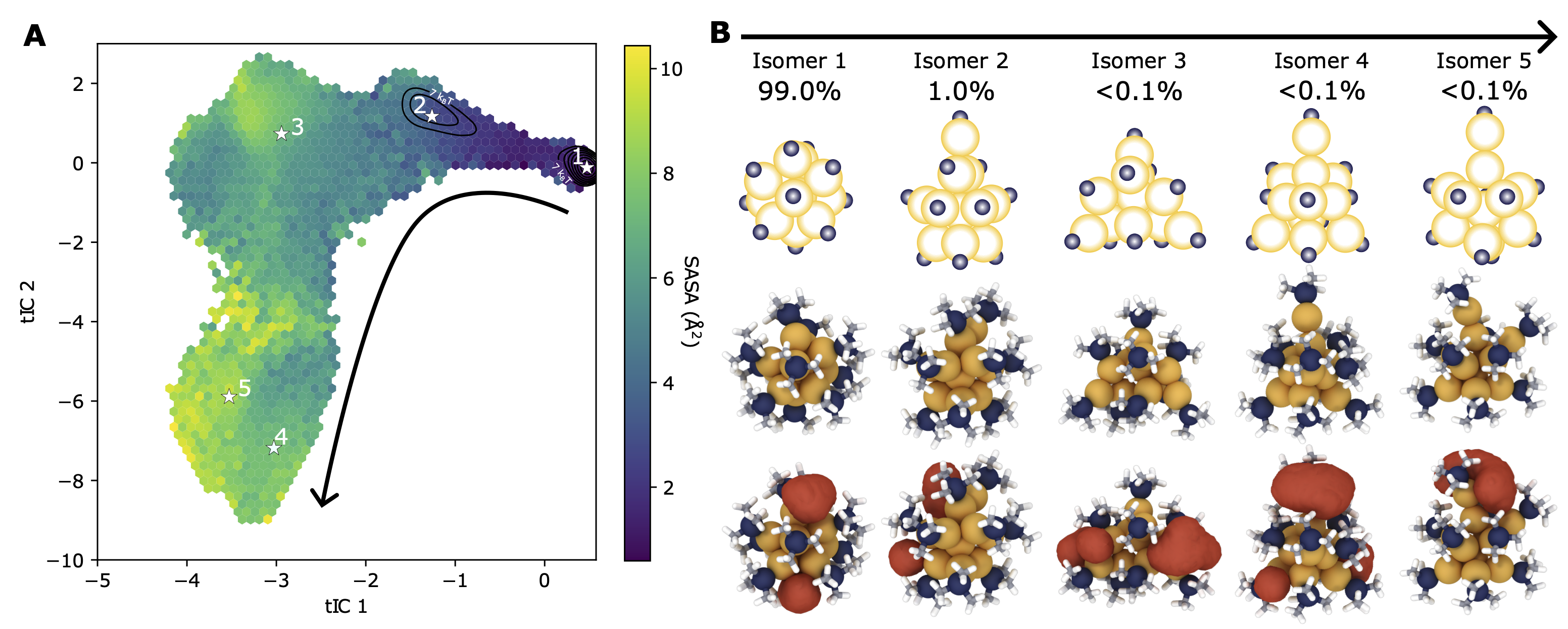}
    \caption{Solvent-accessible surface area (SASA) of the \ce{Au11L10} cluster. A. Free-energy surface in TICA space, colored by the average SASA (Å$^2$) of structures occupying each region. B. Representative isomers observed in the simulation, shown as schematic representations (top), all-atom structures (middle), and including accessible probe positions (bottom). Red spheres denote the centers of 1.4~\AA-radius probes that can contact the Au core while remaining free of steric overlap with the surrounding ligands.}
    \label{fig:sasa}
\end{figure}

To quantify the accessibility of the gold core across the equilibrium ensemble, we calculated the solvent-accessible surface area (SASA)\cite{shrake_environment_1973} of the gold atoms while treating the phosphine ligands as steric barriers, similar to previous work investigating the dynamical properties of thiolate-protected gold clusters.\cite{sabooni_asre_hazer_thermal_2025} SASA was evaluated for every frame of the MD trajectories using the \textit{mdtraj} implementation of the golden section spiral algorithm with a probe radius of 1.4~Å,\cite{mcgibbon_mdtraj_2015} comparable to the size of a water molecule. Only surface area that could be reached by the probe without intersecting any ligand atoms contributed to the reported SASA, providing a simple geometric measure of how accessible the gold core is to potential adsorbates. The relationship between structural dynamics and catalytic accessibility is illustrated for the \ce{Au11L10} cluster in Figure~\ref{fig:sasa}. The dominant equilibrium structure (isomer~1, 99.0\% population) exposes very little of the gold surface, with a typical SASA of only $\sim$1~Å$^2$. In contrast, several minor isomers expose substantially larger regions of the gold core, reaching SASA values up to $\sim$10~Å$^2$. Although these structures are only sparsely populated, they may disproportionately contribute to the catalytic activity by providing access to adsorption sites that are inaccessible in the dominant equilibrium geometry. The MSM relaxation timescales further show that these higher-energy, more surface-exposed isomers are dynamically accessible rather than kinetically trapped, with interconversion between isomer 1 and the remaining isomers occurring on a characteristic timescale of 156~ps.

For other cluster sizes, it is also common that the most stable isomer has a relatively well-protected gold core, but there are some exceptions like \ce{Au4L4}, which prefers an open planar structure over the more protected tetrahedral geometry. A summary of all SASA values projected on the free-energy plots for each cluster size is included in SI Fig. \ref{fig:sasa_summary}. These observations demonstrate the importance of considering dynamic structural ensembles when relating nanocluster structure to function. Evaluating only the most stable geometry can substantially underestimate the accessibility of the gold core and therefore overlook transient but potentially important catalytic conformations. More generally, 
the catalytic properties of ligand-protected nanoclusters should be understood as ensemble-averaged quantities, reflecting both the equilibrium populations of different isomers and their individual structural characteristics rather than those of a single structure.
 
\section*{Conclusion}
We comprehensively characterized the room-temperature equilibrium and dynamic properties of phosphine-stabilized gold nanoclusters by combining the near first-principles accuracy of MLIPs with the long-timescale coarse-grained dynamics provided by MSMs. Rather than viewing atomically precise nanoclusters as single static structures, our results reveal that they are more appropriately described as dynamic ensembles of metastable isomers connected by thermally activated rearrangements. Across the experimentally known cluster sizes investigated here, the equilibrium ensemble often differs substantially from the crystallographic picture obtained by single-crystal X-ray diffraction. In many cases, experimentally determined crystal structures constitute only minor components of the gas-phase equilibrium ensemble, while entropically favored geometries dominate at finite temperature. Notably, we find that multiple crystallographically distinct polymorphs of \ce{Au9(PR3)8} rapidly interconvert within a single metastable state, demonstrating that distinct crystal structures do not necessarily correspond to separate finite-temperature isomers. The present gas-phase calculations therefore establish the intrinsic finite-temperature structural preferences of the isolated clusters, providing a reference against which changes induced by crystal packing and solvation can be disentangled.

In addition to the dynamic nature of phosphine-stabilized gold nanoclusters, a clear relationship emerges between ligand protection, structural dynamics, and functional accessibility.
Increasing ligand coverage systematically reshapes both the free-energy landscape and the kinetics of isomerization, driving the transition from planar to three-dimensional gold cores while substantially accelerating structural interconversion. Intermediate ligand coverages exhibit the greatest structural diversity, whereas fully ligated clusters are dominated by only a few rapidly interconverting three-dimensional structures. Analysis of the solvent-accessible surface area further shows that geometries with the greatest accessibility of the gold core are frequently minor members of the equilibrium ensemble, highlighting a trade-off between thermodynamic stability and accessibility that may be important for catalytic applications. 
In conclusion, ligands are found to govern not only the equilibrium structures of nanoclusters but also their kinetics and the accessibility of reactive sites.

More broadly, this work establishes a general framework for resolving nanocluster free-energy landscapes and isomerization networks directly from molecular dynamics simulations. 
Although developed here for phosphine-stabilized gold nanoclusters, the framework is readily applicable to other classes of ligand-protected nanoclusters. More strongly protected systems, such as gold-thiolate nanoclusters, will likely require more aggressive enhanced-sampling strategies, while solvent effects, ligand exchange, cluster growth, and fragmentation remain beyond the scope of the present model. Explicit treatment of solvation will be particularly important for connecting the gas-phase ensembles reported here to experimentally relevant solution-phase behavior. Overall, this work demonstrates that understanding and ultimately controlling the properties of atomically precise nanoclusters requires moving beyond static crystallographic structures toward a description based on their dynamic structural ensembles.

\section*{Methods}
\phantomsection
\makeatletter
\def\@currentlabelname{Methods}
\makeatother
\label{sec:methods}

\subsection*{Markov State Models for Nanoclusters}

Markov state models describe the dynamics of a molecular system as stochastic transitions between  discrete  states over a lag time $\tau$.\cite{husic_markov_2018,prinz_markov_2011,wang_constructing_2018} Following Koopman theory,\cite{brunton_modern_2022} the dynamics can be represented by a linear propagator acting on state probabilities,
\begin{equation}
\mathbf{p}(t+\tau)=\mathbf{P}^{\top}(\tau)\mathbf{p}(t),
\label{eq:msm}
\end{equation}
where $\mathbf{p}$ is the vector of state populations and $\mathbf{P}$ is the transition probability matrix estimated from the molecular dynamics trajectories. The eigenvectors of $\mathbf{P}$ correspond to the  dynamical modes of the system, while the eigenvalues $\lambda_i$ determine the associated relaxation timescales, $t_i$, of the $i$th slowest mode
\begin{equation}
t_i(\tau)= -\frac{\tau}{\ln |\lambda_i(\tau)|} \quad .
\label{eq:timescale}
\end{equation}
Consequently, constructing an MSM amounts to defining a mapping from the full configuration space to the discrete state space. This is achieved by selecting a structural representation that captures the slow dynamical processes, discretizing the resulting feature space into microstates, and estimating the transition probabilities between those states. An advantage of the MSM formulation is that the transition probabilities can be estimated from a collection of short unbiased simulations, regardless of whether or not those simulations were initialized from the equilibrium distribution.\cite{nuske_markov_2017} To obtain an unbiased estimate of the transition matrix from trajectories that were initiated from selected structures rather than equilibrium configurations, transition counts are reweighted using Koopman reweighting,\cite{wu_variational_2017} as implemented in the deeptime package.\cite{hoffmann_deeptime_2021} This procedure recovers equilibrium populations and transition probabilities from short unbiased trajectories initialized from non-equilibrium configurations.

A key step in constructing  MSMs for nanoclusters is choosing structural descriptors that distinguish different cluster geometries while remaining invariant to rigid translations, rotations, and permutations of equivalent atoms. Here, we employ the same ACE functional form for our structural descriptors as for the MLIP (Figure~\ref{fig:msmfitting}B). The descriptors are comprised of radial and spherical harmonic basis functions with a 15~\AA\ cutoff, sufficiently large to encompass the entire nanocluster. Descriptors are computed for each Au atom in the nanocluster and  are averaged over all Au atoms, yielding rotationally, translationally, and permutationally invariant features for the entire nanocluster. In total, 131 ACE basis functions were used with up to 4-body interactions, including 32 describing Au-only interactions and 99 describing Au-P interactions.

The resulting feature vectors provide a continuous representation of structural evolution throughout the MD trajectories (Figure~\ref{fig:msmfitting}C). The 131 features were projected into a lower dimensional space using  TICA 
(Figure~\ref{fig:msmfitting}D). 
The reduced representation was then partitioned into microstates using K-means++ clustering.\cite{arthur_k-means_2007} MSM fitting hyperparameters, including the TICA lag time (100~fs--10~ps), the number of TICA dimensions (4--8), and the number of cluster centers (600--1200), were selected by maximizing the implied relaxation timescales of the resulting MSMs hierarchically, first optimizing the slowest implied timescale (details in SI Sec.~\ref{subsec:MSMfitting}).

For each simulation frame with Cartesian coordinates $\mathbf{x}_t\in\mathbb{R}^{3N}$, 
the mapping into the discrete state space follows
\begin{equation}
\mathbf{x}_t
\rightarrow
\boldsymbol{\phi}(\mathbf{x}_t)
\rightarrow
\mathbf{f}_t
\rightarrow
s_t,
\end{equation}
where $\boldsymbol{\phi}$ denotes the 131-dimensional ACE feature vector, $\mathbf{f}_t$ is its projection into TICA space, and $s_t$ is the assigned microstate. The resulting discrete state trajectory is used to estimate the transition matrix $\mathbf{P}(\tau)$ in Eq.~\eqref{eq:msm}. Convergence of the implied timescales with increasing MSM lag time was used to assess the validity of the Markovian approximation and the sufficiency of sampling. When convergence was not achieved, additional trajectories were launched from under-sampled regions of configuration space (see SI Sec.~\ref{subsec:MSMfitting}). 

The microstates were subsequently coarse-grained into metastable macrostates  using Perron-Cluster Cluster Analysis (PCCA+),\cite{roblitz_fuzzy_2013} which partitions the microstates according to the slow eigenmodes of the transition matrix (Figure~\ref{fig:msmfitting}E). These eigenmodes identify groups of microstates that rapidly interconvert with one another while exchanging slowly with other groups, thereby defining metastable macrostates (Figure~\ref{fig:msmfitting}F). The number of PCCA macrostates for each system (SI Table~\ref{tbl:fits}) was chosen such that each additional macrostate captured a distinct isomer associated with a slow isomerization process. Structurally distinct configurations that interconverted rapidly were therefore not classified as separate isomers.


\subsection*{Molecular Dynamics Simulations}
\label{subsec:md}

All phosphine-stabilized gold nanoclusters 
were simulated in the gas phase at 300~K using MD as implemented in the LAMMPS simulation package.\cite{thompson_lammps_2022}

It is not guaranteed that room temperature MD simulations initiated from the experimental crystal structures are able to explore all metastable states within the accessible simulation timescales. Replica exchange simulations were therefore used to explore the configurational landscape,\cite{sugita_replica-exchange_1999} from which 32 representative structures spanning the accessible configurations were selected as initial structures for independent room-temperature (NVT) production simulations (for additional details of the replica exchange simulations see SI Sec.~\ref{subsec:simulations}). Since these selected structures were not Boltzmann distributed, the trajectories were reweighted in the construction of the MSM (see previous section) to obtain an unbiased estimate of the coarse-grained dynamics.
The production simulations were run with Langevin dynamics in the NVT ensemble with a damping parameter of 1~ps, a time step of 1~fs/step, and a frame print rate of 100~fs/frame for a total of 288-1536~ns in order to converge the coarse-grained dynamical model for each system (details provided in SI  Sec.~\ref{subsec:MSMfitting}).

Ligand dissociation was rare, occurring in fewer than 1\% of the simulated trajectories. Even so, any trajectory exhibiting ligand dissociation was truncated one frame before the first unbinding event. The resulting MSMs therefore describe only the isomerization dynamics of intact ligand-protected nanoclusters. No ligand migration between gold atoms was observed in the simulations. Accordingly, all reported structural transitions reflect rearrangements of the gold framework rather than changes in ligand connectivity.

\subsection*{Machine-learned interatomic potential}
\label{subsec:mlipfit}
Interatomic interactions between Au, P, C, and H atoms were described using an ACE MLIP\cite{drautz_atomic_2019,lysogorskiy_performant_2021,bochkarev_efficient_2022} which was trained on approximately 42k density functional theory (DFT+D3) calculations of \ce{Au_n(PH3)_m} and \ce{Au_n(P(CH3)3)_m} clusters ($n\leq12$, $m\leq n$). 
The potential considers an interaction radius of up to 7~\AA\ and consists of 5594 basis functions with 9290 trainable parameters. Up to 6-body interactions are included for Au interactions, 5-body interactions for Au-P interactions and all other interactions have up to 4-body interactions. The final architecture is detailed in SI table \ref{tbl:ACEarchitecture}.

An initial dataset for training the potential was sourced from a previous study.~\cite{mccandler_phosphine-stabilized_2023}
This dataset consisted of over 10k DFT geometry relaxations of neutral \ce{Au_n} and \ce{Au_n(PH_3)_m} structures ($n\leq12$, $m\leq n+1$). As geometry relaxation steps can be highly correlated, a maximum of 4 training examples were selected from each geometry relaxation calculation. 
The initial training set contained 19,368 Au-P-H structures. 

Additional training data were generated to extend the potential to trimethylphosphine (\ce{P(CH_3)_3}), which was used throughout this work as a computationally tractable model for experimentally relevant bulky phosphine ligands. This choice was motivated by shortcomings of the simpler \ce{PH3} model, including unrealistically frequent ligand dissociation and \ce{PH3} decomposition into \ce{PH2} + \ce{H}. An initial set of training structures with trimethylphosphine groups was created by swapping the \ce{PH3} ligands in the initial dataset with \ce{P(CH_3)_3} and iteratively rotating these ligands until they were not overlapping (within 2~\AA). A subset of these structures, with a small jitter applied to each atomic position, was added to the training dataset. Additionally, structure snapshots from test simulations as well as isomer geometries that were not in the initial dataset were added to augment the training dataset within an active learning scheme.\cite{lysogorskiy_active_2023} Specifically, the potential was iteratively refined 14 times with an additional 22,796 active learning structures. The resulting MLIP achieves a mean absolute error of 2.14~meV/atom in energies and 43.0 meV/Å in forces on representative structures sampled from the production simulations. Details of the potential architecture, benchmarking, and DFT calculations are provided in SI Sec.~\ref{subsec:mlip}.

\section*{Acknowledgements}
Thanks to Maryam Sabooni Asre Hazer for helpful discussions. D. Sanwal thanks the ICC at the Flatiron Institute for hospitality while a portion of this research was carried out. The Flatiron Institute is a division of the Simons Foundation.

\section*{Supporting Information}
Supporting Information Available: machine learned interatomic potential architecture and benchmarking, PES pre-exploration with parallel tempering, MSM fitting details and MSM results (dynamical maps, isomer geometries, equilibrium populations, and isomerization kinetics) for all cluster sizes, and fitted ACE Au-P-C-H potential.
\newpage

\bibliographystyle{achemso_jabbrv}
\bibliography{aupch}

\end{document}


\setcounter{table}{0}
\setcounter{figure}{0}
\renewcommand{\thefigure}{S\arabic{figure}}
\renewcommand{\theequation}{S\arabic{equation}}
\renewcommand{\thetable}{S\arabic{table}}
\setcounter{equation}{0}
\setcounter{table}{0}
\setcounter{section}{0}
\renewcommand\thesection{\Alph{section}}
\renewcommand\thesubsection{\thesection.\arabic{subsection}}

\title{Supplementary Information: Dynamic Ensembles of Phosphine-Stabilized Gold Nanoclusters}
\author{Caitlin A. McCandler}
\address{Initiative for Computational Catalysis, Flatiron Institute, NY 10010, USA}
\author{Disha Sanwal}
\address{University of Chicago, IL 60637, USA}
\author{Jutta Rogal}
\address{Initiative for Computational Catalysis, Flatiron Institute, NY 10010, USA}

\date{\today}
\maketitle 

\section{Machine learned interatomic potential: architecture, reference calculations and benchmarking}
\label{subsec:mlip}
\subsection{Potential architecture}

\begin{table}[!htb]
     \begin{center}
      \caption{Architecture of final ACE potential. The $n_\text{max}$ and $l_\text{max}$ are reported for each multi-body interaction type (i.e. 2-body/3-body/4-body)}
     \begin{tabular}{ c c c c}
    Elements &  Max Body-Order & $n_{max}$ & $l_{max}$ \\
    \hline
    Au &  6-body & 11/7/3/2/1 & 0/3/3/2/1\\										
    Au-P &  5-body & 11/6/2/1 & 0/3/2/1\\
    All other &  4-body & 11/5/2 & 0/3/1\\
    \hline
     
      \end{tabular}
      \label{tbl:ACEarchitecture}
      \end{center}
      \end{table}

\subsection{Density Functional Theory Calculations}
The energies and forces for the MLIP training data were calculated with D3-corrected DFT at the same level of theory as the initial training dataset.\cite{mccandler_phosphine-stabilized_2023} Calculations were performed using the Vienna Ab-initio Simulation Package (VASP) with a plane-wave basis set and spin-polarization.\cite{kresse_efficient_1996} The exchange and correlation energies were calculated using the Perdew-Burke-Ernzerhof (PBE) form of the generalized gradient approximation (GGA).\cite{perdew_generalized_1996} All clusters were isolated from their periodic image by at least 10~\AA\ of vacuum spacing and were calculated with one k-point, i.e., the $\Gamma$ point. Gaussian smearing was applied with a width of 0.2 eV, a cutoff energy of 520 eV was applied for the plane wave basis set and the electron–ion interactions were described by the projector augmented wave (PAW) method.\cite{kresse_ultrasoft_1999} D3 dispersion energies and forces were added as a correction to the DFT calculated energies and forces.\cite{grimme_consistent_2010} 

\subsection{Benchmarking}
Benchmarking statistics for the MLIP are included in Table~\ref{tbl:errors}, where specific attention is paid to representing the lowest energy structures well. The set of structures that are within 1~eV/atom of the energy hull are labeled as the ``(low)" structures, and these corresponded to mainly structures with trimethylphosphine ligands and more reasonable structures that are most similar to the simulation structures. In the final stage of fitting the potential, structures were selected from testing 300~K simulations and recalculated with DFT to give an estimate of the error of representative simulation structures (labeled dynamic ground states). These structures were not included in the training set but yet have an MAE of 2.14 meV/atom in energies and 43.0 meV/Å in forces. The standard deviations of the energies and forces for each of these testing and training datasets are also included to show that a wide variety of structures were used to fit the potential, not just ground state structures, so that hopefully high energy transition states are also well represented. 

This potential should not be used for unbound ligands or inter-cluster interactions, which were not included in the training set. Ligand dissociation events in these simulations were very rare, only occurring in less than 1\% of the trajectories. All simulations where ligands unbind were truncated a frame before the unbinding event to avoid fitting the dynamical models to different system sizes. 

MD simulations using the ACE potential were performed in parallel, with 32 independent simulations running concurrently on each CPU node. The average simulation speeds were 22.6 ns/day for the \ce{Au4L4} cluster and 8.0 ns/day for the \ce{Au11L10} cluster, corresponding to overall throughputs of 723.2 and 256.0 ns/day per CPU node, respectively.

\begin{figure}[!htb]
    \centering
    \includegraphics[width=\linewidth]{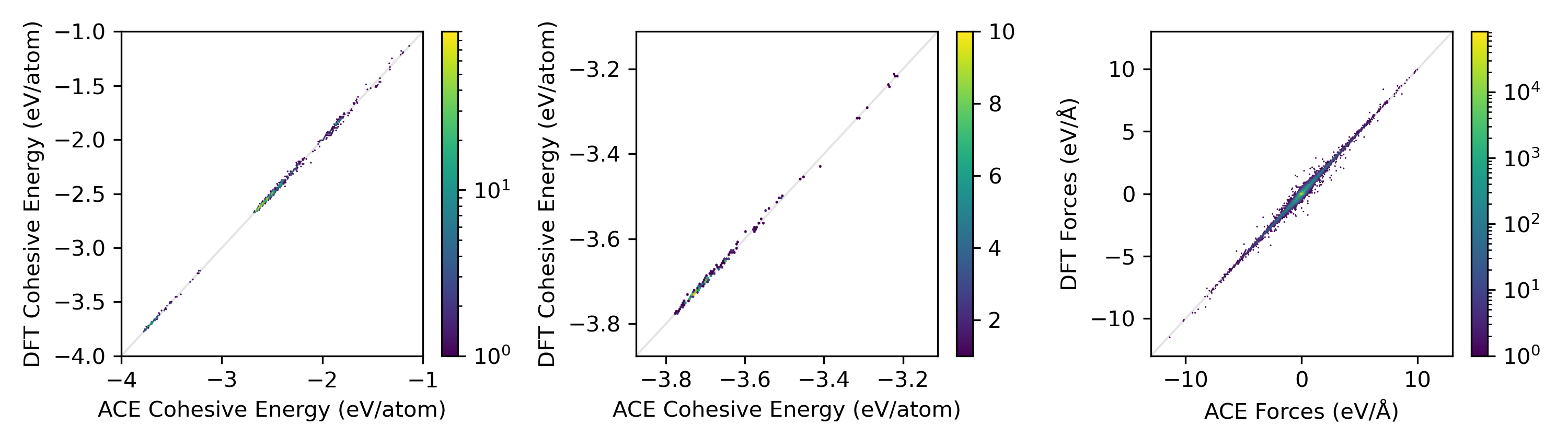}
    \caption{DFT vs. ACE energies for (left) test dataset, (center) energies for test (low) dataset, and (right)forces for test dataset}
    \label{fig:benchmark_parity}
\end{figure}

\begin{table}[]
\renewcommand{\arraystretch}{0.5}
\caption{Benchmarking statistics for the fitted Au-P-C-H ACE potential on the testing/training data, on the testing/training data within 1eV/atom of the energy hull ``(low)", as well as on a selection of snapshots representing each of the isomers from the simulations.}
\begin{tabular}{c|ccc|ccc|}
                     & \multicolumn{3}{c|}{Energy (meV/atom)}                                           & \multicolumn{3}{c|}{Forces (meV/Å)}                                              \\
Dataset              & RMSE                              & MAE                              & STD       & RMSE                              & MAE                              & STD       \\ \hline
Test (low)           & \multicolumn{1}{c|}{10.13 (4.33)} & \multicolumn{1}{c|}{6.64 (3.03)} & 439 (103) & \multicolumn{1}{c|}{68.2 (68.0)} & \multicolumn{1}{c|}{41.8 (39.0)} & 4854 (9866) \\
Train (low)          & \multicolumn{1}{c|}{10.10 (5.00)} & \multicolumn{1}{c|}{6.55 (3.29)} & 440 (117) & \multicolumn{1}{c|}{66.8 (67.6)} & \multicolumn{1}{c|}{41.3 (38.1)} & 4510 (9277) \\
Dynamic ground states & \multicolumn{1}{c|}{2.74}             & \multicolumn{1}{c|}{2.14}            & 10.95          & \multicolumn{1}{c|}{68.3}             & \multicolumn{1}{c|}{43.0}            &  1773        
\end{tabular}
\label{tbl:errors}
\end{table}

\null
\newpage
\null
\newpage

\section{PES pre-exploration with replica exchange}
\label{subsec:simulations}
Replica exchange was performed as an initial step before starting production simulations to discover unknown energy minima in each cluster's potential energy surface (PES)\cite{sugita_replica-exchange_1999}. Initial geometries for this pre-exploration were sourced from crystal structures as well as the dataset used to fit the MLIP, where all ligands were replaced with trimethylphosphine. Each geometry was simulated for 1~ns with 4 replicas held at 300~K, 333~K, 366~K, and 400~K, with an attempt at a tempering swap every 1000 steps (1~ps), allowing for frequent temperature exchanges between images. 
\begin{figure}[!htb]
    \centering
    \includegraphics[width=.5\linewidth]{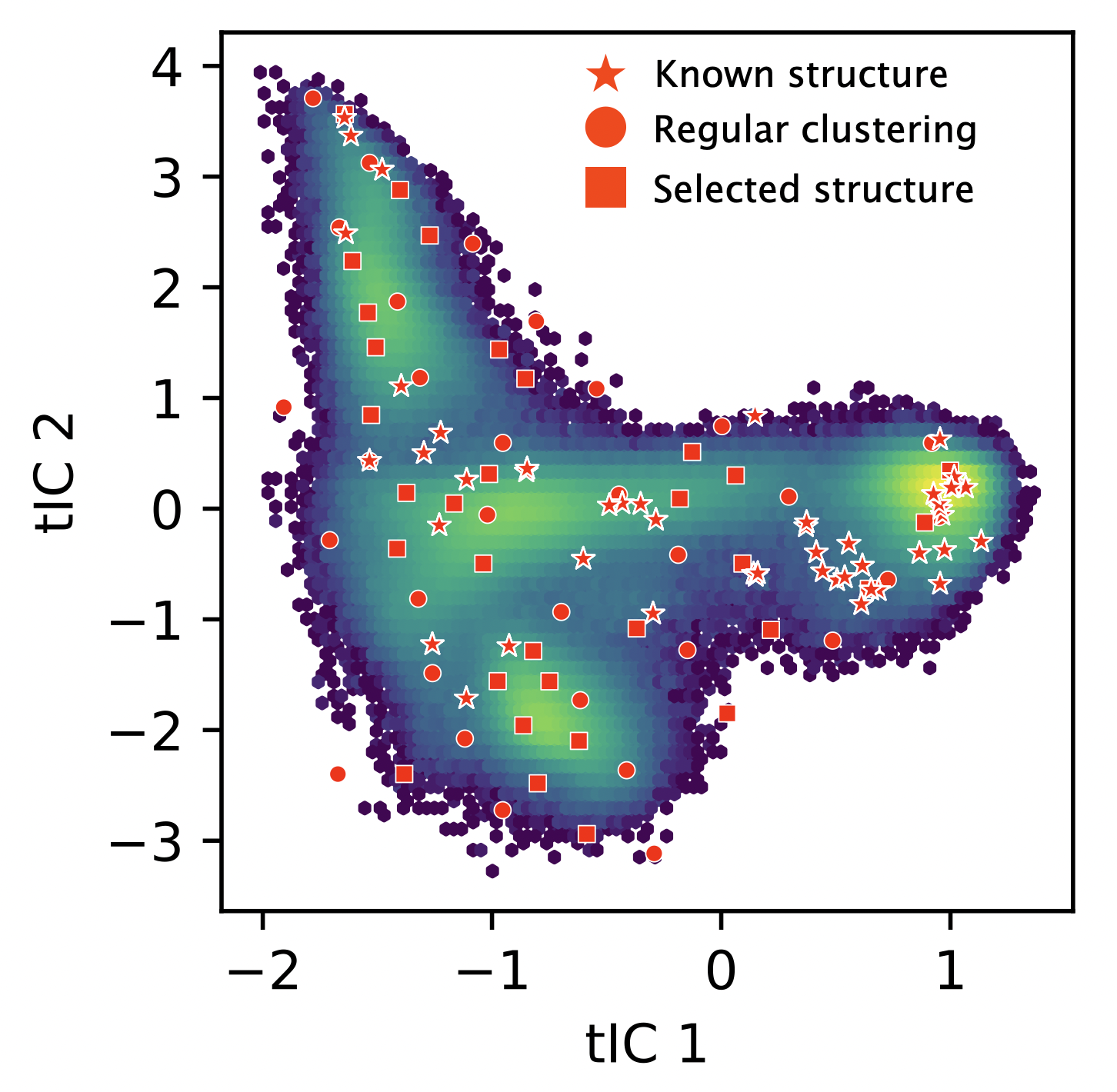}
    \caption{Two-dimensional TICA projection of the replica exchange trajectories (hexbin density map) used for selecting initial configurations for the production simulations. Red stars indicate the known starting structures used to initialize the pre-exploration, red circles denote the 32 regular-space clustering centers, and red squares show the randomly selected production starting structures, with one structure chosen from each cluster. The selected configurations span the configurational space sampled during the replica exchange simulations.}
    \label{fig:preexploration}
\end{figure}
Sampling at elevated temperatures promoted barrier crossing and more extensive exploration of the PES, reducing the likelihood that subsequent 300 K production simulations would remain trapped in local minima. Initial configurations for the production simulations were selected by applying regular-space clustering (32 centers) to the explored configurational space and selecting one structure from each cluster (32 structures). Clustering was performed in a two-dimensional TICA space constructed from the replica exchange trajectories using a lag time of 500 fs (Fig.~\ref{fig:preexploration}) and the structure selection was random.

\null
\newpage

\section{Markov state models: fitting and dynamical maps}
\label{subsec:MSMfitting}
\subsection{Hyperparameter selection}
MSMs were fitted for each system by optimizing the model hyperparameters: the TICA lag time, the number of TICA dimensions retained and the number of K-means++ cluster centers. For a well-constructed MSM, the implied timescales of the slow dynamical modes should become independent of the chosen MSM lag time. At the same time, the MSM lag time should remain shorter than the implied timescale of the corresponding process to avoid overestimating kinetic separation, and longer converged implied timescales generally indicate models that better capture the underlying slow dynamics.\cite{arbon_markov_2024} The model hyperparameters were optimized hierarchically. First, the combination yielding the largest converged implied timescale for the slowest dynamical mode was identified. Hyperparameter sets producing values within 1\% of this maximum were kept, and this procedure was repeated sequentially for the second-slowest mode, third-slowest mode, and so forth until an optimal set of hyperparameters was chosen. For each process, the MSM lag time was selected as the largest value in the set \{500~fs, 1~ps, 2.5~ps, 5~ps, 10~ps, 25~ps, 50~ps, 100~ps, 250~ps, 500~ps\} that remained below the implied timescale of that process. The final hyperparameter values are listed in Table~\ref{tbl:fits}.

\begin{table}[b]
\caption{Final hyperparameter choices for MSM fittings for each \ce{Au_{n}L_{m}} cluster and the resulting time scales of the slowest dynamical modes observed in each system.}
\renewcommand{\arraystretch}{0.5}
\begin{tabular}{c|c|c|c|c|c|c}
System  & Macrostates & \begin{tabular}[c]{@{}c@{}}TICA\\ dimensions\end{tabular} & \begin{tabular}[c]{@{}c@{}}TICA\\ time lag (ps)\end{tabular} & \begin{tabular}[c]{@{}c@{}}Cluster\\ centers\end{tabular} & \begin{tabular}[c]{@{}c@{}}Slowest\\ timescale (ps)\end{tabular} & \begin{tabular}[c]{@{}c@{}}Simulation\\ length (ns)\end{tabular} \\ \hline
\ce{Au4L4}   & 2           & 5                                                         & 0.5                                                          & 2000                                                      & 5                                                                 & 288\\
\ce{Au6L6}   & 5           & 5                                                         & 1                                                            & 2000                                                      & 97                                                                & 288\\
\ce{Au7L6}   & 6           & 4                                                         & 0.5                                                          & 1400                                                      & 1693                                                              & 336\\
\ce{Au7L7}   & 4           & 5                                                         & 5                                                            & 2000                                                      & 153                                                               & 336\\
\ce{Au8L0}   & 4           & 7                                                         & 1                                                            & 2000                                                      & 11656                                                             & 672\\
\ce{Au8L2}   & 7           & 7                                                         & 1                                                            & 600                                                       & 10253                                                             & 1344\\
\ce{Au8L4}   & 16          & 8                                                         & 1                                                            & 1400                                                      & 27280                                                             & 1152\\
\ce{Au8L6}   & 20          & 8                                                         & 0.5                                                          & 1000                                                      & 1731                                                              & 1536\\
\ce{Au8L7}   & 6           & 8                                                         & 1                                                            & 2000                                                      & 648                                                               & 528\\
\ce{Au8L8}   & 4           & 4                                                         & 5                                                            & 2000                                                      & 655                                                               & 288\\
\ce{Au9L8}   & 5           & 7                                                         & 1                                                            & 2000                                                      & 743                                                               & 432\\
\ce{Au11L10} & 5           & 8                                                         & 0.1                                                          & 1400                                                      & 156                                                               & 864 \end{tabular}
\label{tbl:fits}
\end{table}

\subsection{MSM convergence}
\begin{figure}[!htb]
    \centering
    \includegraphics[width=\linewidth]{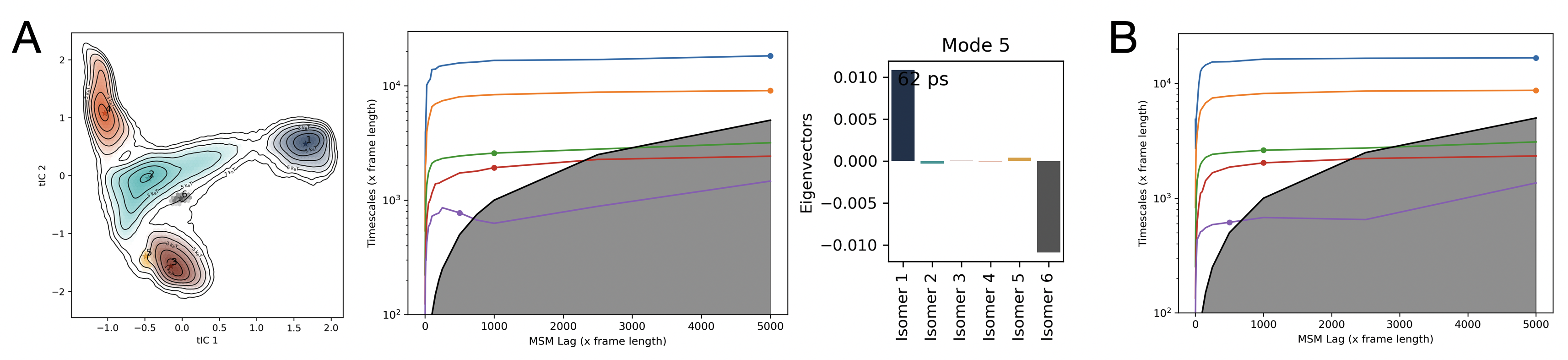}
    \caption{The Markov state model for the \ce{Au7L6} cluster was initially (A) unconverged in the 5th dynamical mode (corresponding to the interconversion of isomer 1 and isomer 6),  but after adding 3 rounds of 1 ns simulations from 16 initial structures identified as being close to the transition state for this unconverged mode, (B) the mode converged.}
    \label{fig:refining_modes}
\end{figure}

The MSMs for each cluster system were iteratively refined by identifying unconverged dynamical modes and increasing sampling in those regions. Figure \ref{fig:refining_modes} shows a representative example of this process. In this example, the 5th slowest mode in the \ce{Au7L6} system was not well sampled, as evidenced by its timescale remaining dependent on the MSM lag time. Often the timescales that are the most difficult to sufficiently sample are those that are very slow or are between rare states. In this case, the poorly sampled mode corresponded to transitions between isomers 1 and 6, where isomer 6 is only rarely observed. To improve sampling of unconverged modes, new trajectories were initiated from previously sampled frames located near transitions between the states contributing to the mode (i.e. transitioning between relevant states either in the next step or in the previous step). When a mode involved multiple isomers, starting structures were selected from transitions spanning all relevant state pairs identified by the mode. This process was repeated for all systems until the timescales converged as a function of the MSM lag time. The total simulation time for each system to achieve convergence was 288 ns, 288~ns, 336~ns, 336~ns, 1536~ns, 528~ns, 288~ns, 432~ns, and 864~ns of production simulations for \ce{Au4L4}, \ce{Au6L6}, \ce{Au7L6}, \ce{Au7L7}, \ce{Au8L6}, \ce{Au8L7}, \ce{Au8L8}, \ce{Au9L8}, and \ce{Au11L10}, respectively.

Convergence was also assessed by monitoring the reweighted free-energy landscapes as additional trajectories were incorporated. Sampling was considered sufficient once the free-energy basins, their relative populations, and the overall landscape remained unchanged with further increases in simulation data.

\subsection{MSMs for each \ce{Au_{n}L_{m}} cluster}

\begin{figure}[!htb]
    \centering
    \includegraphics[width=\linewidth]{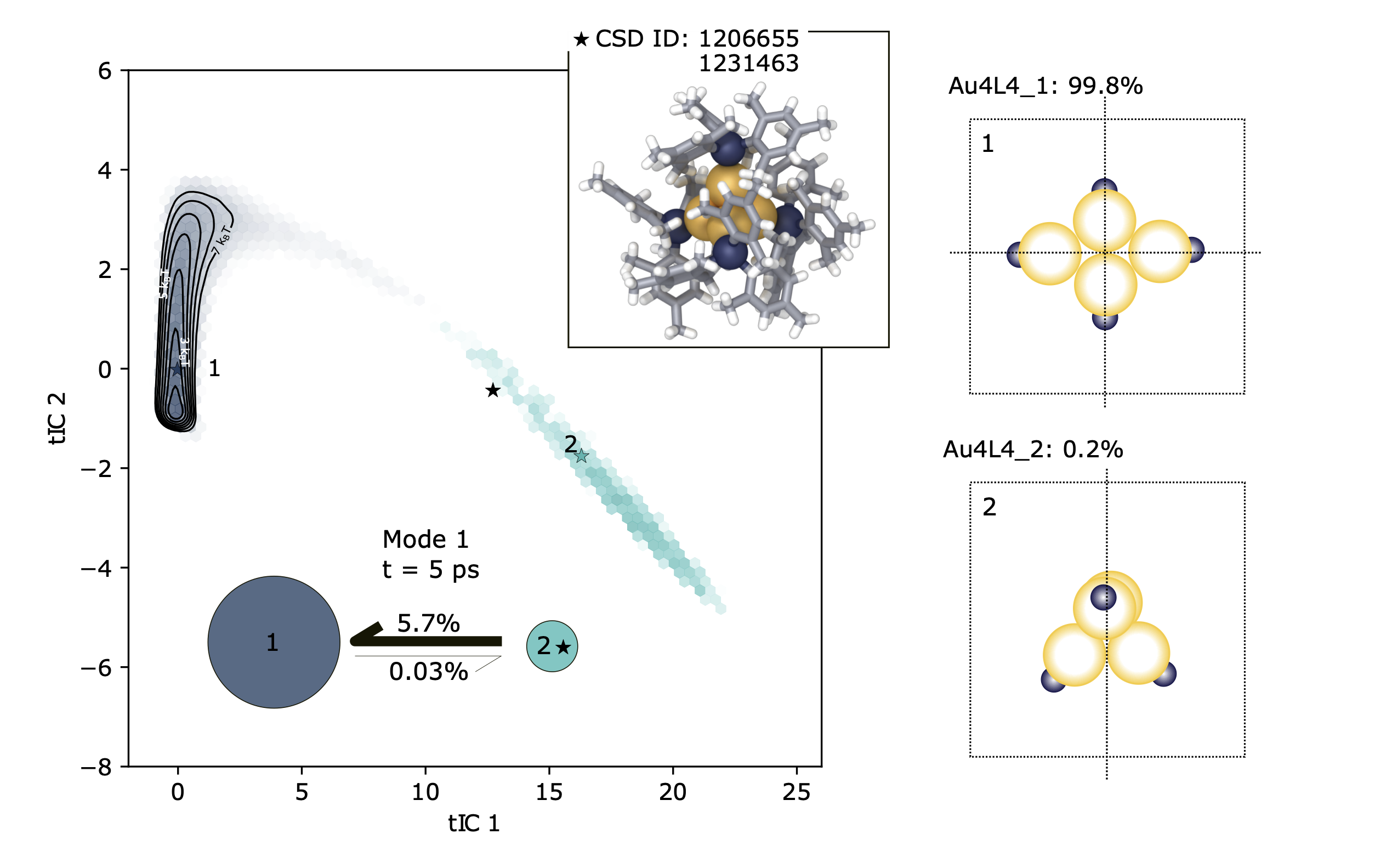}
    \caption{Dynamical map, known crystal structure (CSD IDs: 1206655, 1231463), coarse grained model of transition flux ($\tau=$~100 fs), dynamical modes, and isomer schematics and equilibrium populations of the \ce{Au4L4} (L=trimethylphosphine) cluster at 300~K in the gas phase. The crystal structure corresponds to a minor isomer (Isomer 2, 0.2\%).}
    \label{fig:Au4L4}
\end{figure}

\begin{figure}[!htb]
    \centering
    \includegraphics[width=\linewidth]{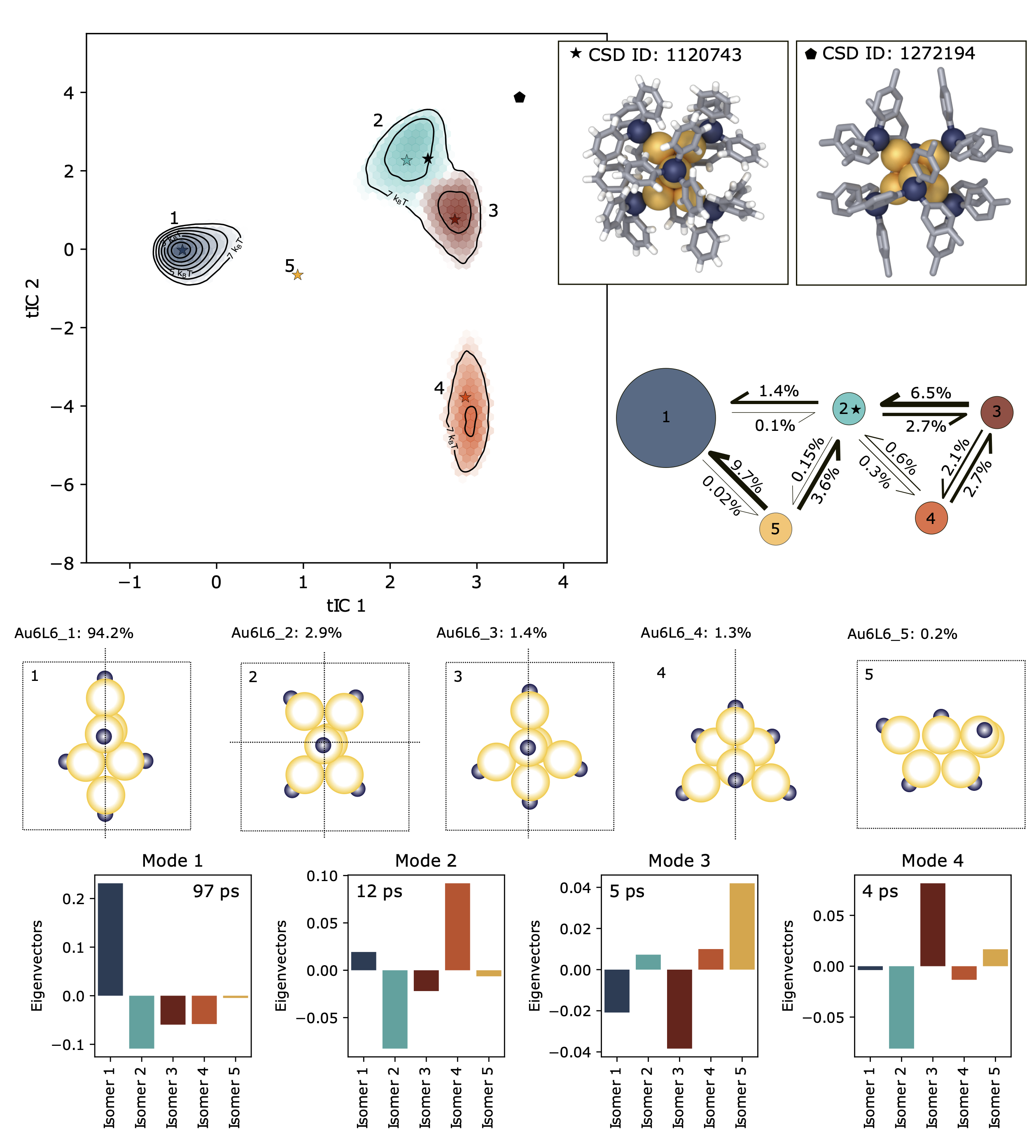}
    \caption{Dynamical map, known crystal structure (CSD IDs: 1120743, 1272194), coarse grained model of transition flux ($\tau=$~100 fs), dynamical modes, and isomer schematics and equilibrium populations of the \ce{Au6L6} (L=trimethylphosphine) cluster at 300~K in the gas phase. One crystal structure (CSD ID: 1120743) corresponds to a minor isomer (Isomer 2, 2.9\%), and the other does not correspond to any isomers.}
    \label{fig:Au6L6}
\end{figure}

\begin{figure}[!htb]
    \centering
    \includegraphics[width=\linewidth]{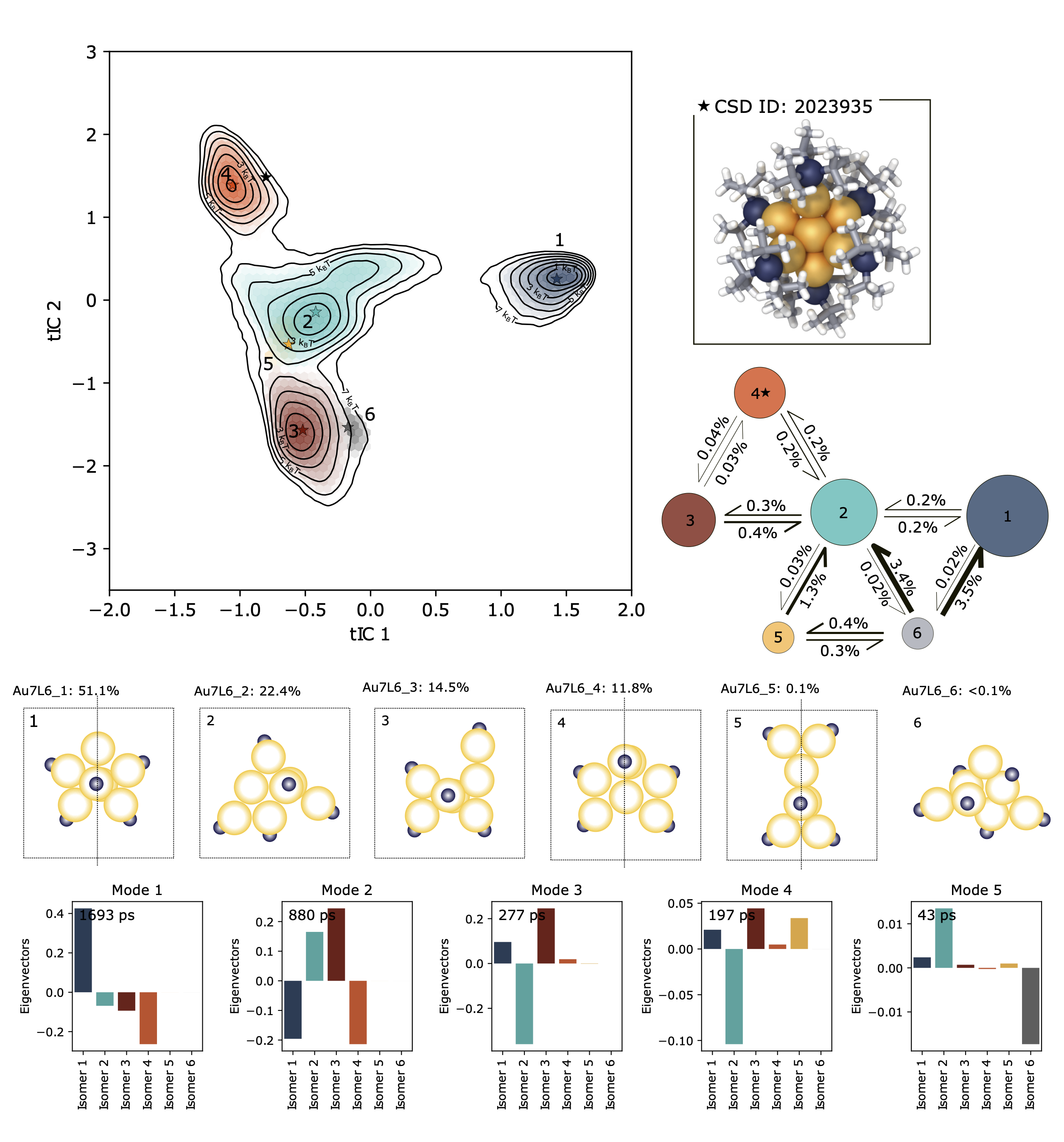}
    \caption{Dynamical map, known crystal structure (CSD ID: 2023935), coarse grained model of transition flux ($\tau=$~100 fs), dynamical modes, and isomer schematics and equilibrium populations of the \ce{Au7L6} (L=trimethylphosphine) cluster at 300~K in the gas phase. The crystal structure corresponds a slight distortion from a minor isomer (Isomer 4, 11.8\%).}
    \label{fig:Au7L6}
\end{figure}

\begin{figure}[!htb]
    \centering
    \includegraphics[width=\linewidth]{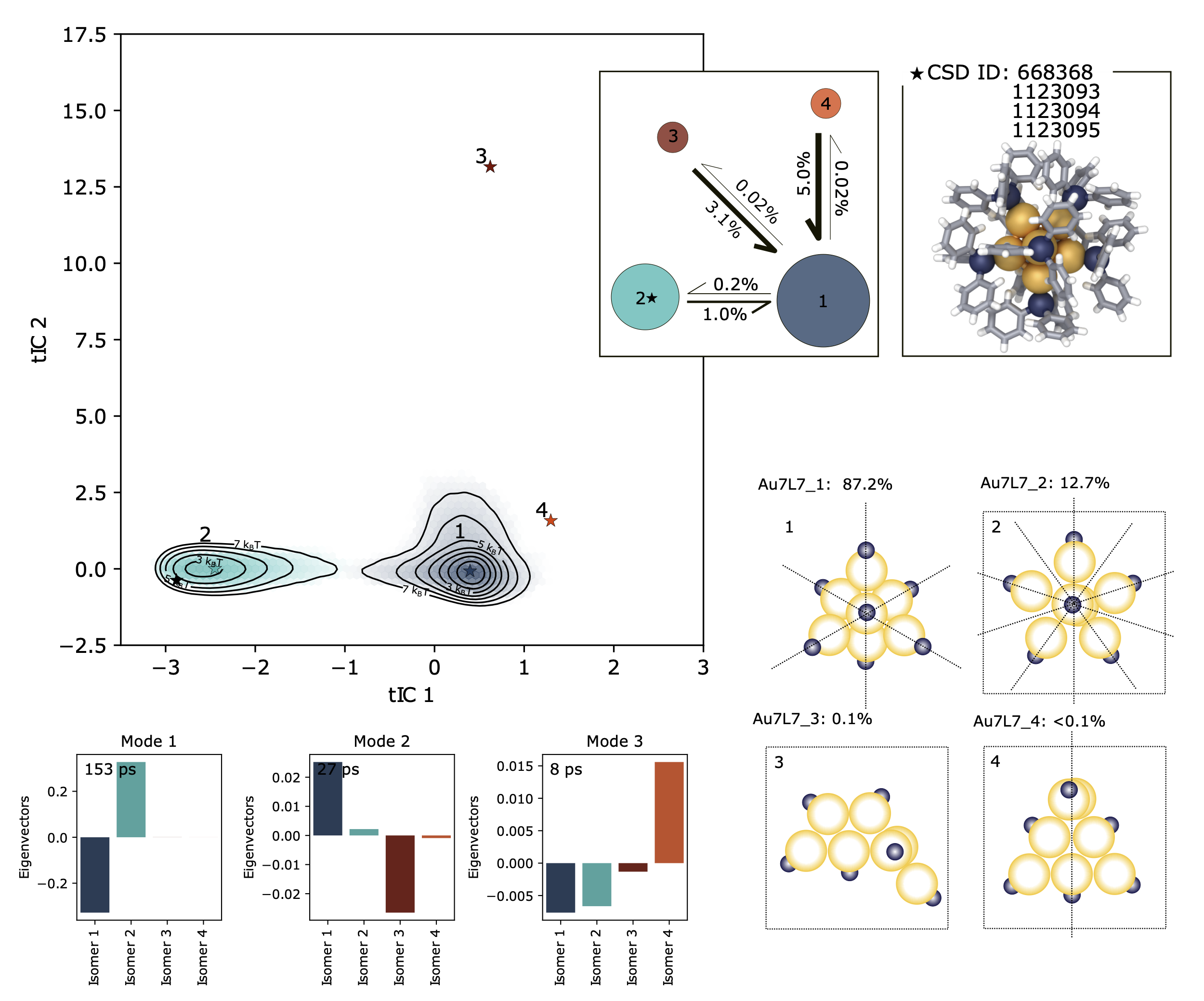}
    \caption{Dynamical map, known crystal structure (CSD IDs: 668368, 1123093, 1123094, 1123095), coarse grained model of transition flux ($\tau=$~100 fs), dynamical modes, and isomer schematics and equilibrium populations of the \ce{Au7L7} (L=trimethylphosphine) cluster at 300~K in the gas phase. The crystal structure corresponds a minor isomer (Isomer 2, 12.7\%).}
    \label{fig:Au7L7}
\end{figure}

\begin{figure}[!htb]
    \centering
    \includegraphics[width=\linewidth]{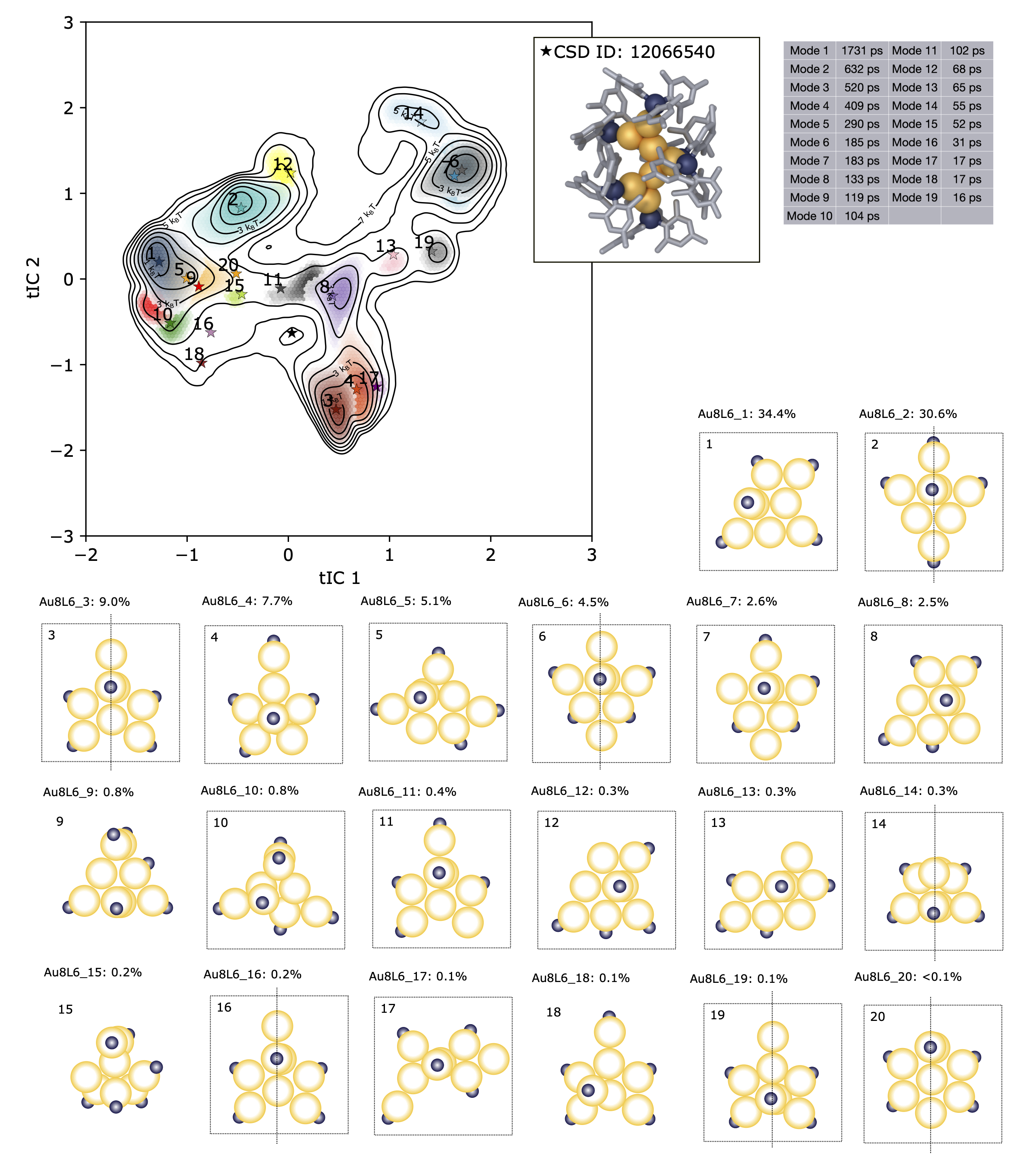}
    \caption{Dynamical map, known crystal structure (CSD ID: 2023935), timescales of dynamical modes, and isomer schematics and equilibrium populations of the \ce{Au8L6} (L=trimethylphosphine) cluster at 300~K in the gas phase. The crystal structure does not correspond to any metastable isomers.}
    \label{fig:Au8L6}
\end{figure}

\begin{figure}[!htb]
    \centering
    \includegraphics[width=\linewidth]{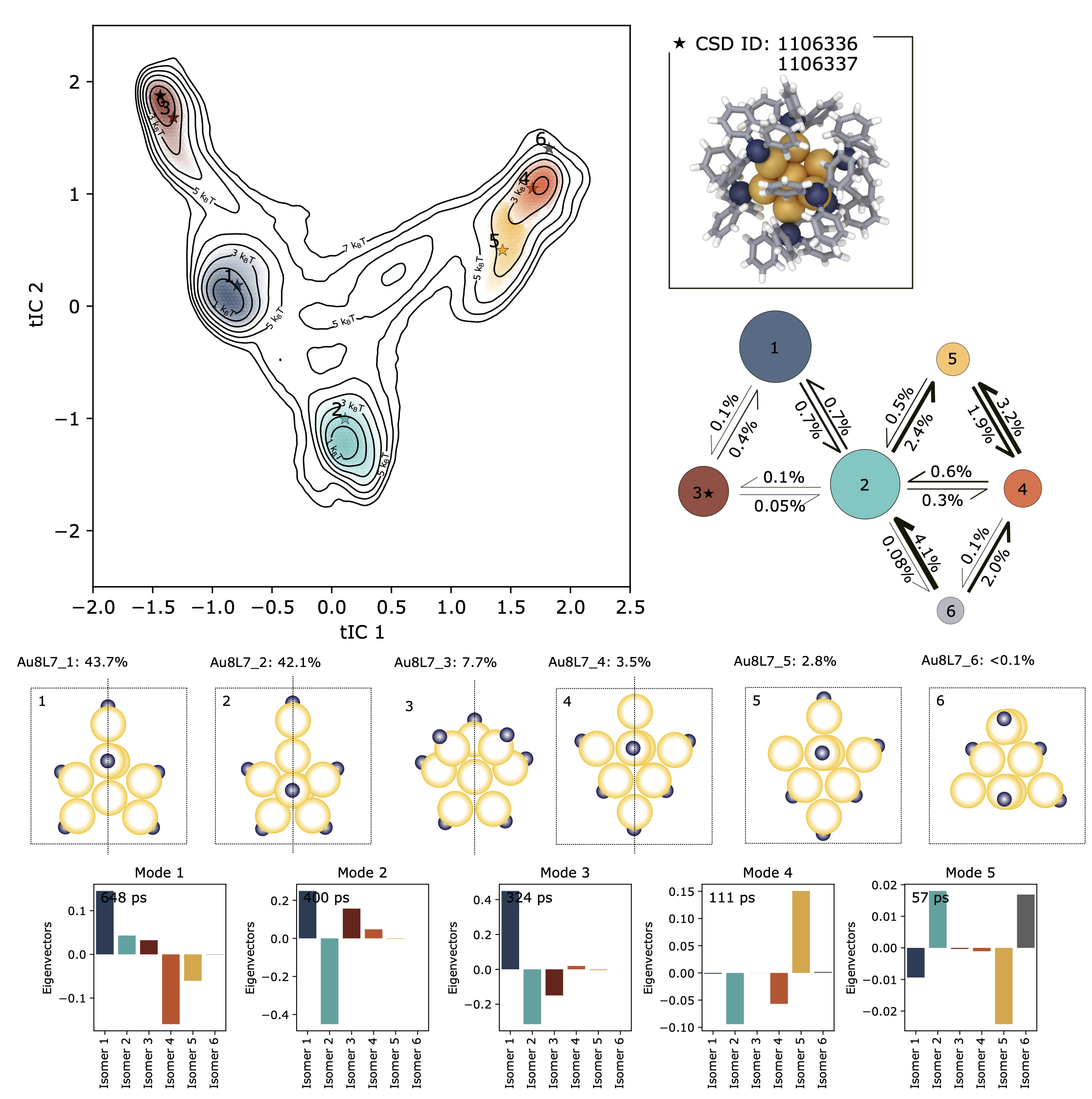}
    \caption{Dynamical map, known crystal structure (CSD IDs: 1106336, 1106337), coarse grained model of transition flux ($\tau=$~100 fs), dynamical modes, and isomer schematics and equilibrium populations of the \ce{Au8L7} (L=trimethylphosphine) cluster at 300~K in the gas phase. The crystal structure corresponds a minor isomer (Isomer 3, 7.7\%).}
    \label{fig:Au8L7}
\end{figure}

\begin{figure}[!htb]
    \centering
    \includegraphics[width=\linewidth]{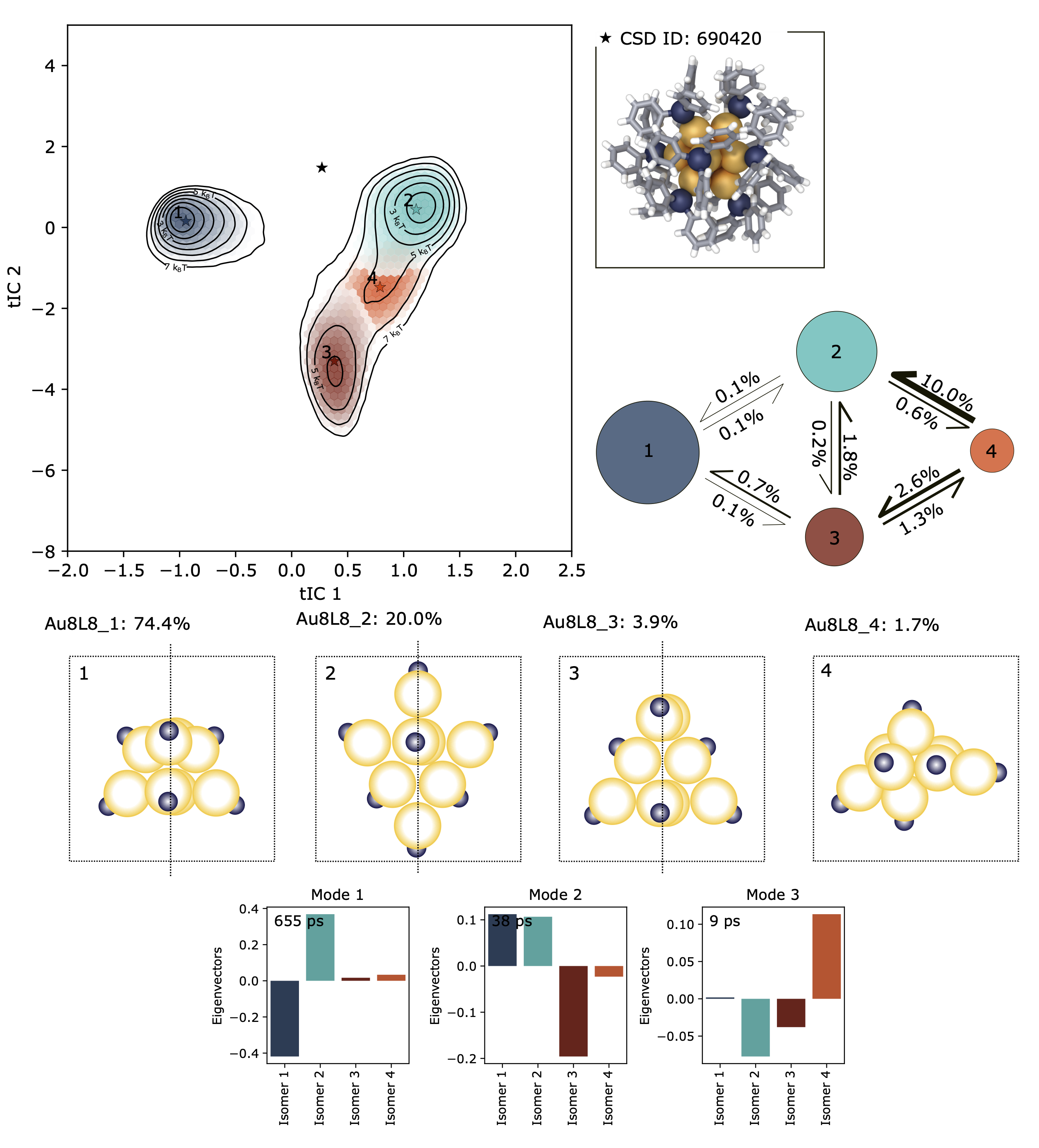}
    \caption{Dynamical map, known crystal structure (CSD ID: 690420), coarse grained model of transition flux ($\tau=$~100 fs), dynamical modes, and isomer schematics and equilibrium populations of the \ce{Au8L8} (L=trimethylphosphine) cluster at 300~K in the gas phase. The crystal structure does not correspond to any metastable isomers.}
    \label{fig:Au8L8}
\end{figure}

\begin{figure}[!htb]
    \centering
    \includegraphics[width=\linewidth]{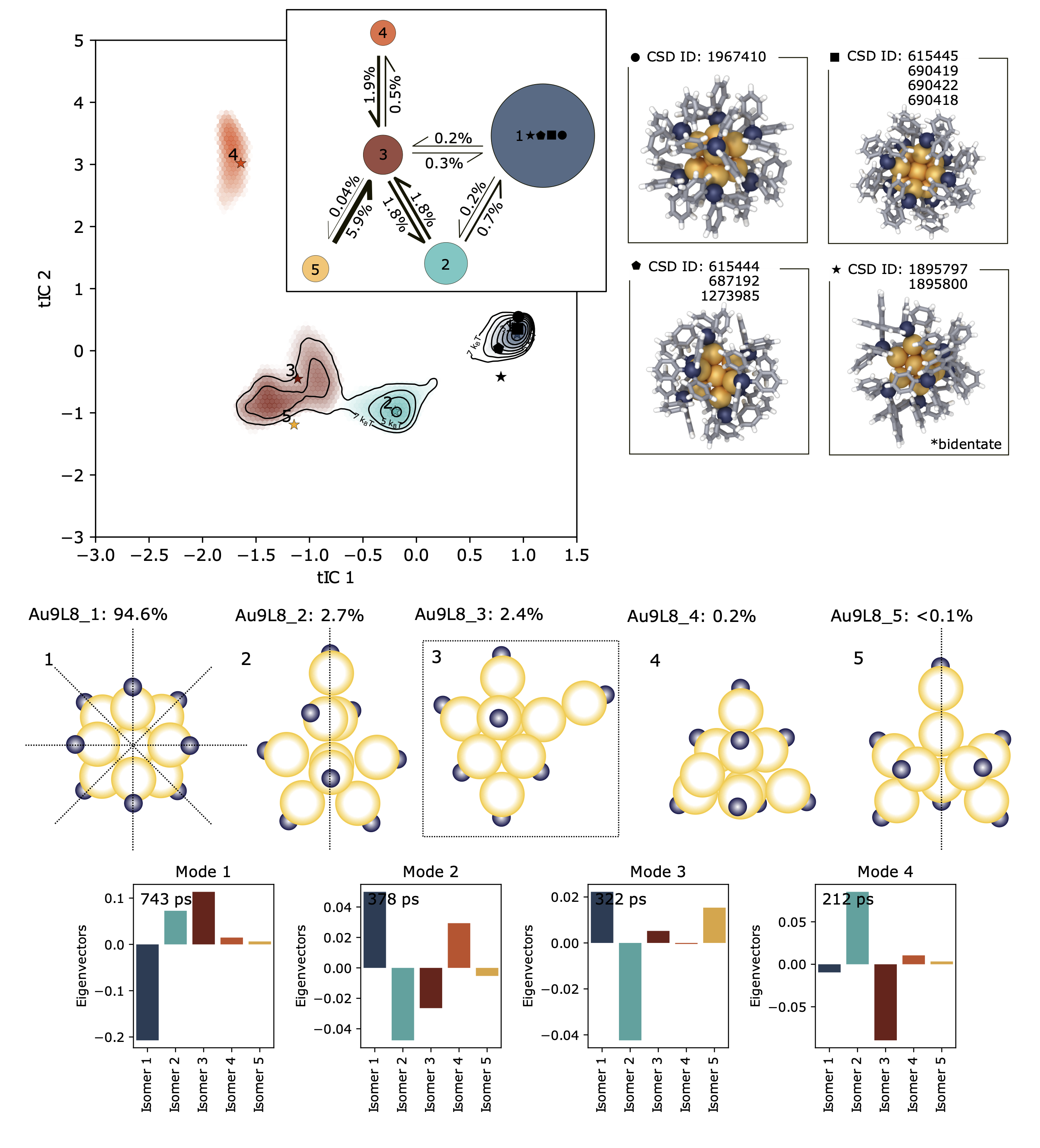}
    \caption{Dynamical map, known crystal structures (CSD IDs: 1967410, 615445, 690419, 690422, 690418, 615444, 687192, 1273985, 1895797, 1895800), coarse grained model of transition flux ($\tau=$~100 fs), dynamical modes, and isomer schematics and equilibrium populations of the \ce{Au9L8} (L=trimethylphosphine) cluster at 300~K in the gas phase. All of the crystal structures correspond to a major isomer (Isomer 1, 94.6\%).}
    \label{fig:Au9L8}
\end{figure}

\begin{figure}[!htb]
    \centering
    \includegraphics[width=\linewidth]{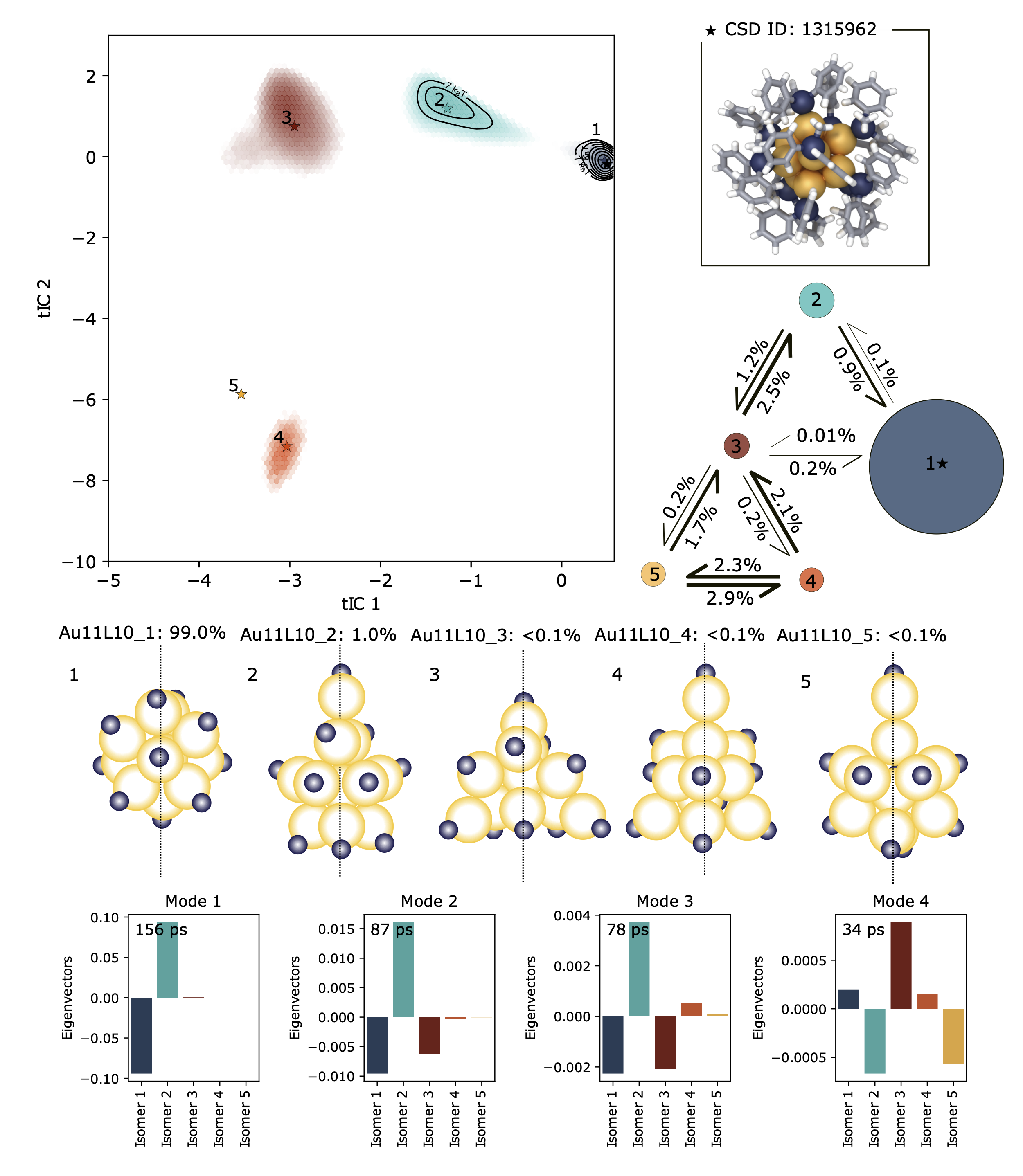}
    \caption{Dynamical map, known crystal structure (CSD ID: 1315962), coarse grained model of transition flux ($\tau=$~100 fs), dynamical modes, and isomer schematics and equilibrium populations of the \ce{Au11L10} (L=trimethylphosphine) cluster at 300~K in the gas phase. The crystal structure corresponds a major isomer (Isomer 1, 99.0\%).}
    \label{fig:Au11L10}
\end{figure}

\FloatBarrier
\subsection{Additional details of \ce{Au_{8}L_{m}} clusters}
\begin{figure}[!htb]
    \centering
    \includegraphics[width=\linewidth]{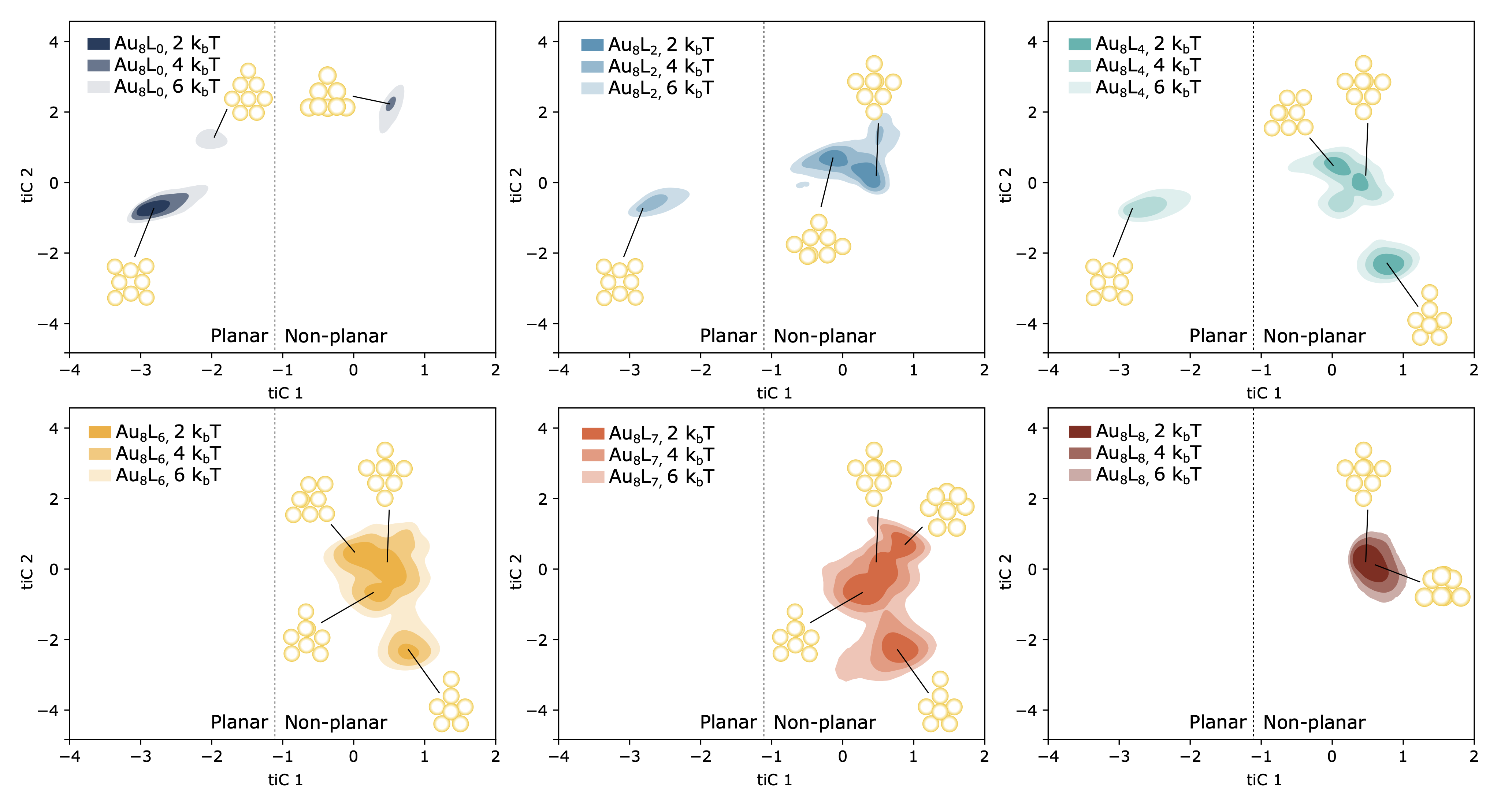}
    \caption{Supplement to Figure~3 showing the individual dynamical maps of each \ce{Au8L_{m}} (L=~trimethylphosphine, m=~0, 2, 4, 6, 7, 8) cluster isomerization at 300~K in the gas phase. Free energy isolines are projected into a TICA space that is fitted only to the geometry of the gold core (i.e. only Au-Au interactions are considered).}
    \label{fig:au8breakout}
\end{figure}
\begin{figure}[!htb]
    \centering
    \includegraphics[width=\linewidth]{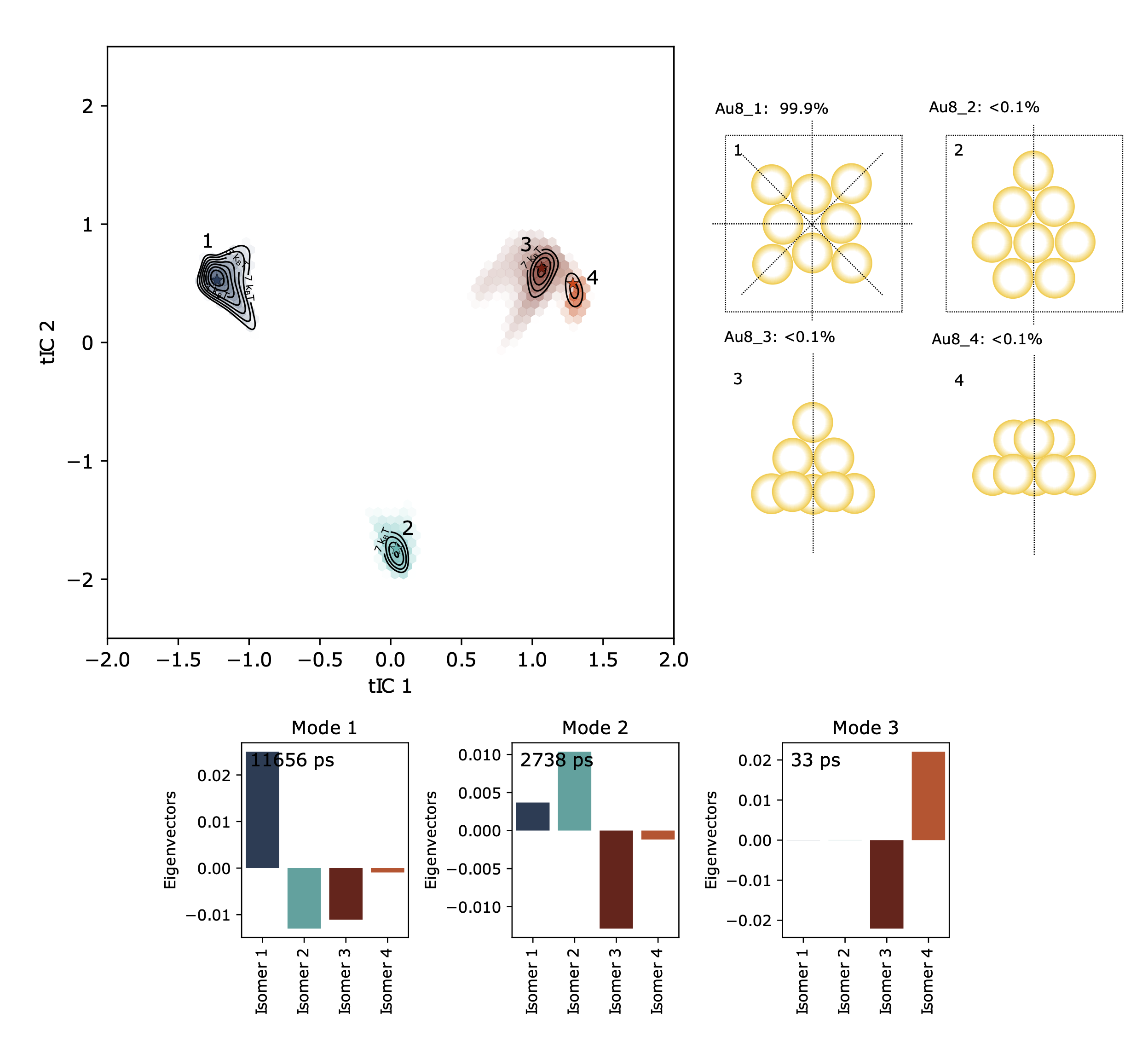}
    \caption{Dynamical map, dynamical modes, and isomer schematics and equilibrium populations of the \ce{Au8} cluster at 300~K in the gas phase.}
    \label{fig:Au8L0}
\end{figure}

\begin{figure}[!htb]
    \centering
    \includegraphics[width=\linewidth]{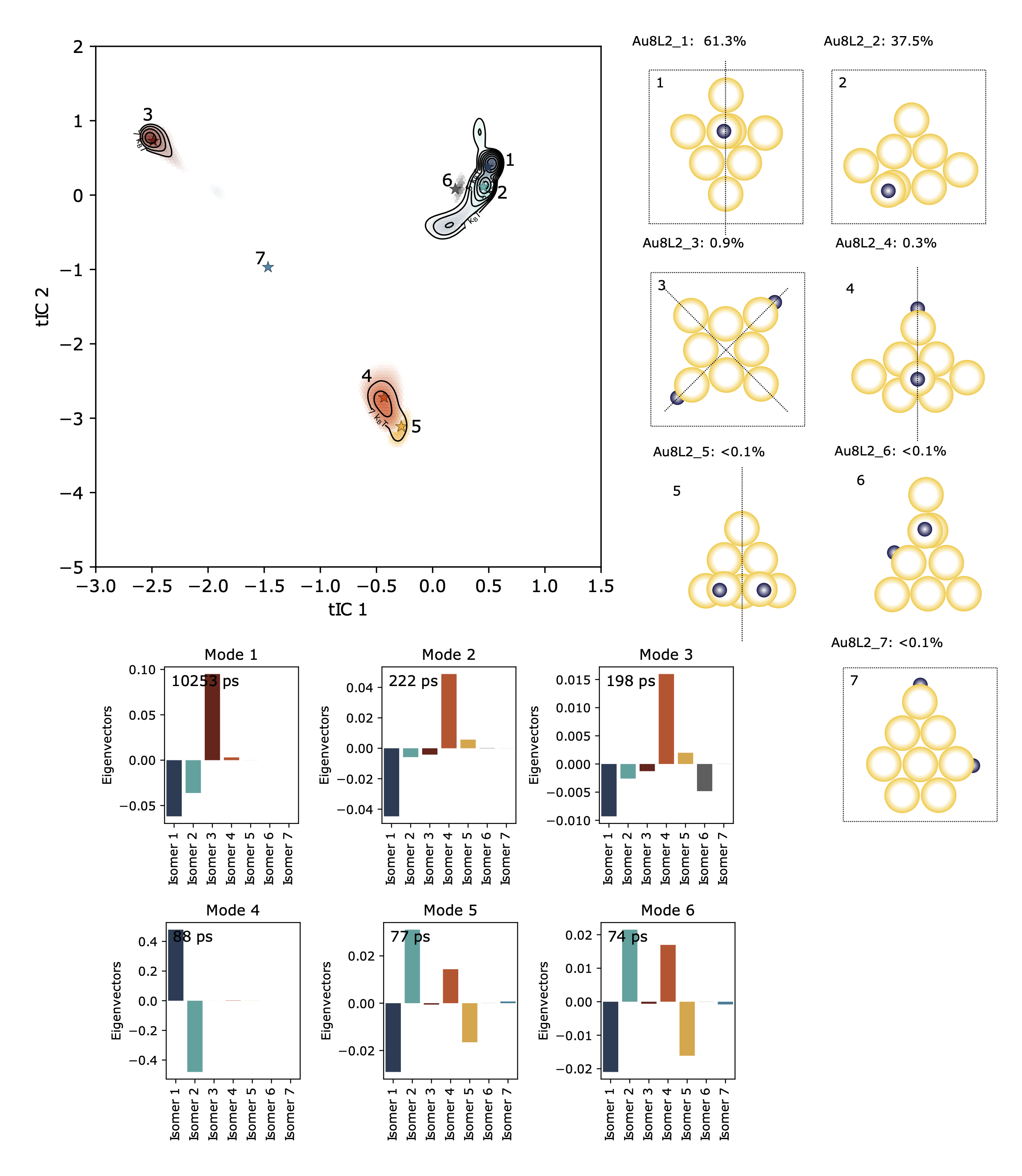}
    \caption{Dynamical map, dynamical modes, and isomer schematics and equilibrium populations of the \ce{Au8L2} (L=trimethylphosphine) cluster at 300~K in the gas phase.}
    \label{fig:Au8L2}
\end{figure}

\begin{figure}[!htb]
    \centering
    \includegraphics[width=\linewidth]{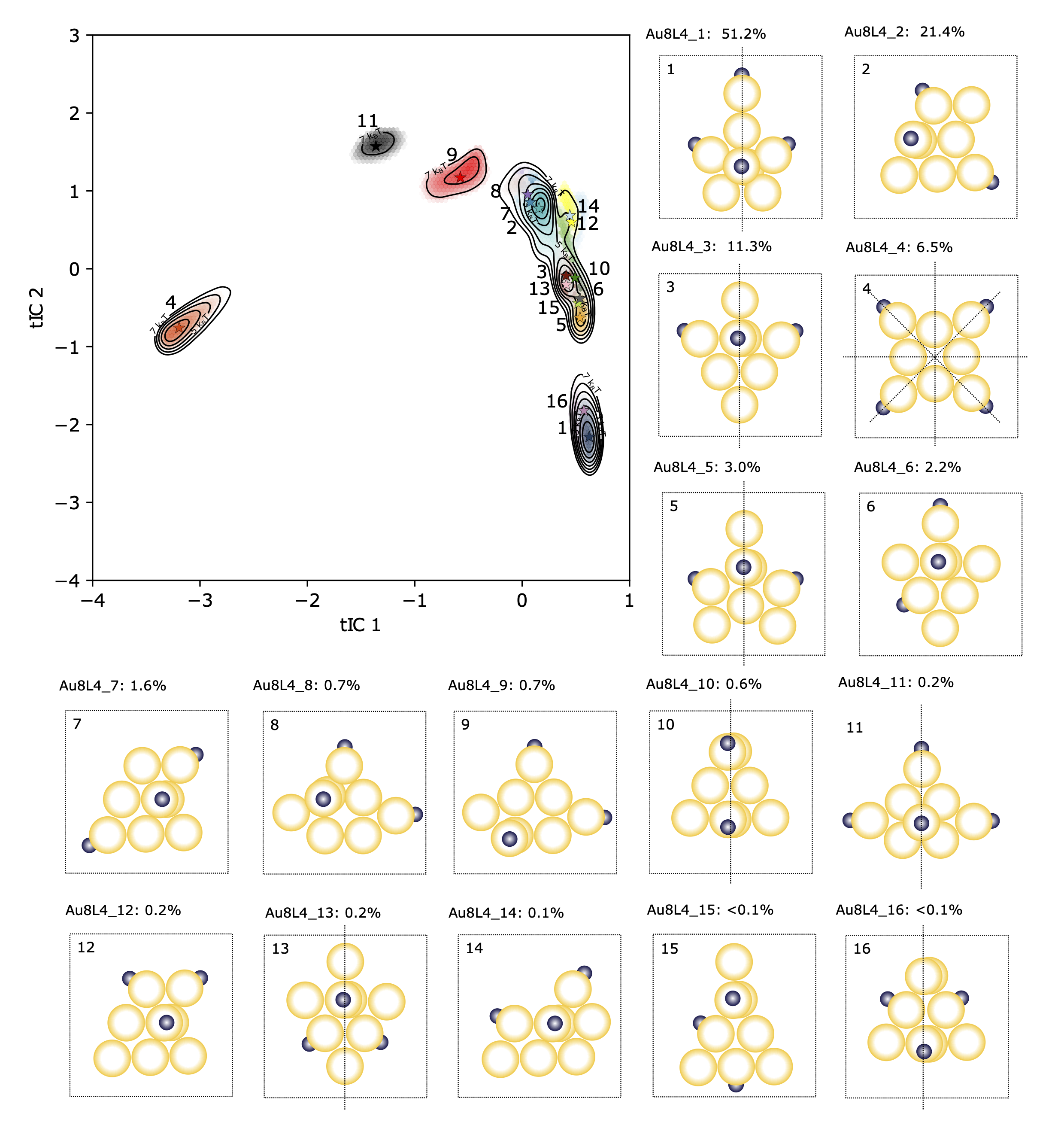}
    \caption{Dynamical map and isomer schematics and equilibrium populations of the \ce{Au8L4} (L=trimethylphosphine) cluster at 300~K in the gas phase.}
    \label{fig:Au8L4}
\end{figure}

\FloatBarrier

\subsection{SASA maps}

\begin{figure}[!htb]
    \centering
    \includegraphics[width=\linewidth]{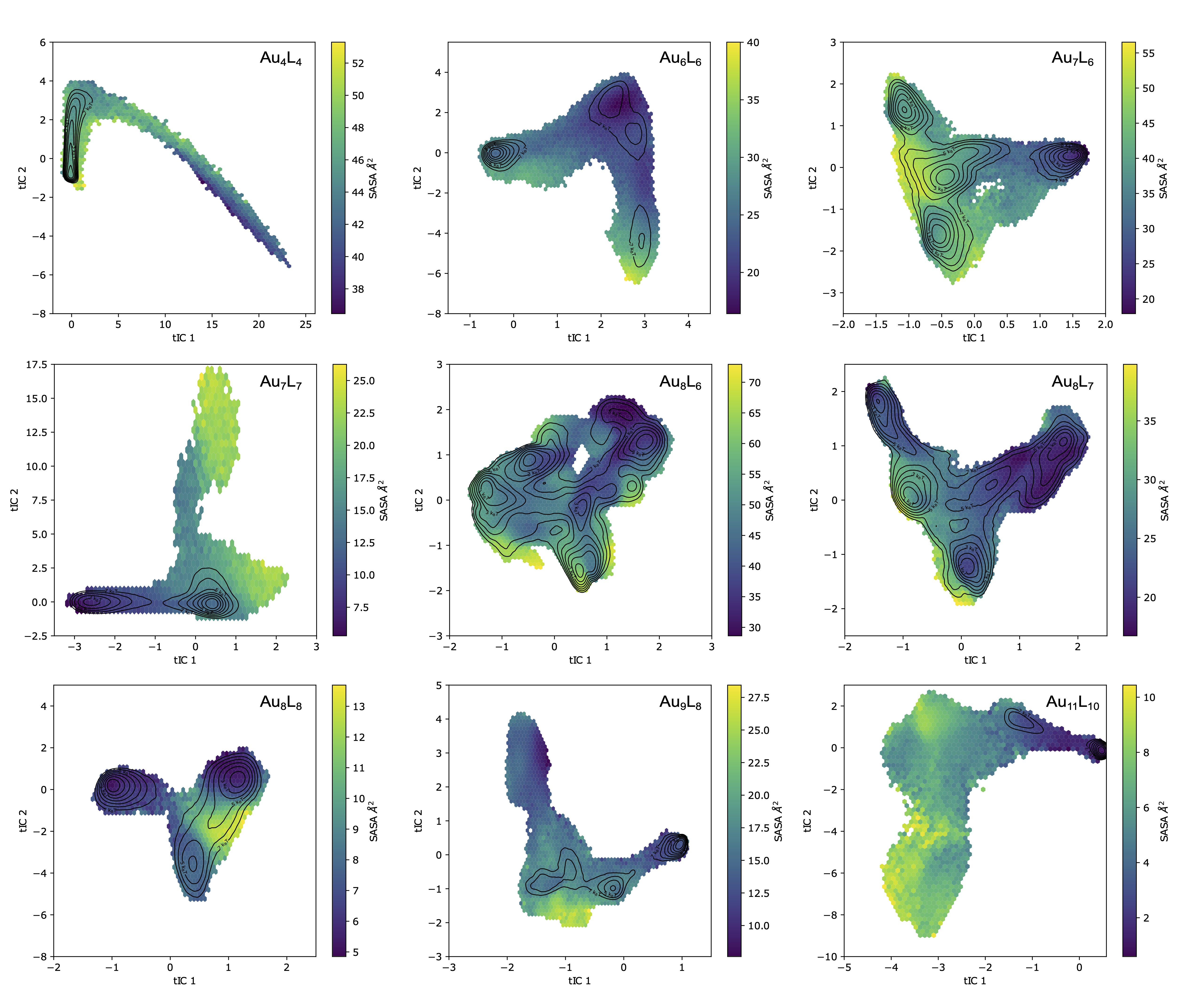}
    \caption{Solvent accessible surface area (SASA) and free energy isolines projected into TICA space for each cluster \ce{Au_{n}L_{m}} (L=~trimethylphosphine) at 300~K in the gas phase.}
    \label{fig:sasa_summary}
\end{figure}
\newpage

\bibliographystyle{achemso_jabbrv}
\bibliography{aupch2}